\documentclass[12pt,a4paper]{article}
\usepackage[english]{babel}
\usepackage[utf8]{inputenc}
\usepackage[T1]{fontenc}
\usepackage{amsmath}
\usepackage{amsfonts}
\usepackage{amssymb}
\usepackage[colorinlistoftodos, color=green!40, prependcaption]{todonotes}
\usepackage{authblk}
\usepackage{cite}
\usepackage{amsthm}
\usepackage{mathtools}
\usepackage{physics}
\usepackage{xcolor}
\usepackage{graphicx}
\usepackage[left=23mm,right=13mm,top=35mm,columnsep=15pt]{geometry} 
\usepackage{adjustbox}
\usepackage{placeins}
\usepackage[T1]{fontenc}
\usepackage{lipsum}
\usepackage{csquotes}

\usepackage[linktoc=page]{hyperref}
\hypersetup{
    pdfauthor={F. Ambrosino, A. Duci, P. A. Grassi, and S. Penati},
    colorlinks=true, citecolor=blue, urlcolor=blue,
    pdfstartview={FitH}, 
pdftitle={SuSymTFT}, 
linkcolor=blue, 
citecolor=blue, 
filecolor=magenta, 
}
\begin{document}

\title{SymTFT in Superspace}

\author[1]{Federico Ambrosino\thanks{\texttt{federicoambrosino25@gmail.com}}}
\author[2]{Alessandro Duci\thanks{\texttt{a.duci@campus.unimib.it}}}
\author[3]{Pietro Antonio Grassi\thanks{\texttt{pietro.grassi@uniupo.it}}}
\author[2]{Silvia Penati\thanks{\texttt{silvia.penati@mib.infn.it}}}

\affil[1]{\normalsize \textit{Perimeter Institute for Theoretical Physics, Waterloo, Ontario N2L 2Y5, Canada}}
\affil[2]{\normalsize \textit{Dipartimento di Fisica, Universit\`a degli Studi di Milano--Bicocca and INFN, Sezione di Milano--Bicocca, Piazza della Scienza 3, 20126 Milano, Italy}}
\affil[3]{\normalsize \textit{Universit\`a del Piemonte Orientale, viale T.~Michel, 11, 15121 Alessandria, Italy}\\\normalsize \textit{INFN, Sez.~Torino, via P.~Giuria, 1, 10125, Torino, Italy}}

\date{\today} 

\renewcommand{\thefootnote}{\fnsymbol{footnote}}
\maketitle
\setcounter{footnote}{0}
\renewcommand{\thefootnote}{\arabic{footnote}}

\begin{abstract}
We propose a manifestly supersymmetric formulation of the Symmetry Topological Field Theory (SuSymTFT) for theories with supersymmetry.  The SymTFT is a framework that helps organizing symmetries and anomalies of a QFT. Albeit a lot of activity in the field has been devoted to the construction of the SymTFT for the bosonic symmetry structure, the fermionic case has not been analyzed in detail. Here, we consider the most prominent example of  theories exhibiting fermionic symmetries, that is supersymmetric models. These are most naturally formulated on supermanifolds, where the supergeometry approach allows for a manifest organization of the symmetries according to their fermionic grading. We provide the general construction of the SuSymTFT as a super-BF theory living in a (n|m)-dimensional supermanifold and check our proposal in two particular examples, the compact and the chiral super-bosons in two dimensions. 
\end{abstract}

\clearpage
\tableofcontents
\clearpage
\section{Introduction}
The modern notion of symmetry in Quantum Field Theory (QFT) goes well beyond ordinary unitary group-like operators  acting on local operators. The seminal paper \cite{Gaiotto:2014kfa} has led to a true revolution in our understanding of symmetries: Symmetries of QFTs have been naturally associated to topological defects, allowing for a vast generalization of their notion, including higher-form symmetries, non-invertible symmetries and, more generally, categorical (non-group-like) structures. For recent reviews on these developments cfr.\  \cite{Schafer-Nameki:2023jdn, Brennan:2023mmt, Bhardwaj:2023kri, Shao:2023gho,Luo:2023ive,Iqbal:2024pee}. 
A particularly efficient way of packaging these data is through the symmetry topological field theory (SymTFT) \cite{Felder:1999mq,Fuchs:2002cm,Kong:2014qka,Gaiotto:2020iye,Apruzzi:2021nmk,Freed:2022qnc,Kong:2020cie,
Kaidi:2022cpf,
Bhardwaj:2023ayw,
Carqueville:2023jhb} . This is  a topological theory in one dimension higher which encodes generalized symmetries, their defects and their 't Hooft anomalies. The advantage of the SymTFT is that it provides a unified framework to treat charged and charge operators on the same footing, while also serving as an organizing tool to handle their anomalies. Another advantage is that it allows to neatly describe topological manipulations of the symmetries - such as gauging and stacking with invertible phases - in terms of simple operations on the bulk theory.

In its most developed form, this viewpoint was first understood for finite or discrete symmetry data 
(cfr.\ e.g.\ \cite{Freed:2024TopologicalSymmetry}), and was then realized in many geometric and string-theoretic constructions \cite{Apruzzi:2022HigherForm6d,Apruzzi:2023StringTheorySymTFT,vanBeest:2023MTheorySymTFT,Kaidi:2023NonInvertibleSymTFT}. More recently, the SymTFT paradigm has been extended to continuous internal symmetries, first in the abelian case and then for continuous non-abelian groups \cite{Brennan:2024ContinuousSymTFT,Antinucci:2024zjp,Bonetti:2025NonAbelianSymTFT}, and very recently to continuous spacetime and flavor symmetries \cite{Apruzzi:2025SpacetimeSymTFT, Jia:2025jmn}. For broad reviews of generalized symmetries and SymTFT, see for instance \cite{SchaferNameki:2024ICTP,Bhardwaj:2024Lectures,Luo:2024LectureNotes}.

Most of this story, however, has been developed for bosonic symmetry structures, while their fermionic analogues  have remained comparably unexplored. Fermionic generalized symmetries exist, and their investigation has recently started \cite{Wang:2023FermionicHigherForm, Ambrosino:2024FermionicGeneralized, Grassi:2025tfp}. The most prominent example of a fermionic (spacetime) symmetry is surely supersymmetry. These new developments strongly suggest that the SymTFT paradigm should admit an intrinsically supersymmetric formulation.\\
This expectation arises  naturally within a geometric approach. 
Supersymmetric field theories are most naturally formulated in supermanifolds (cfr.\ \cite{Castellani:2015ata} and references therein).  In superspace, supersymmetry acts geometrically along the odd directions,  and bosonic and fermionic currents are assembled into supermultiplets. In this language, the relation between conserved currents, background fields, anomalies and defects can often be expressed more uniformly than in their components.  Reformulating SymTFT directly in superspace is therefore not only aesthetically natural, but also conceptually useful. It is the setting in which bosonic and fermionic symmetries can be treated on equal footing from the outset.  

In this paper we develop such a formulation. The main tools that we use come from the geometry of supermanifolds and their integration theory, sometimes referred to as supergeometry \cite{Witten:2012bg,Castellani:2015ata,Castellani:2015dis}. 
Given a supersymmetric theory living on a supermanifold $\mathcal{SM}^{(d|m)}$, we construct a supersymmetric SymTFT, or \textbf{SuSymTFT}, defined on a bulk supermanifold $\mathcal{SM}^{(d+1|n)}$ whose boundary contains the physical superspace. The odd bulk dimension $n$ is determined by the bulk supersymmetry and by the amount preserved at the boundary; in the standard half-BPS situation one expects $n=2m$. The resulting bulk topological theory is designed to encode the full symmetry sector of the boundary model, including bosonic and fermionic symmetries, as well as the corresponding anomaly inflow. In the bosonic case, moving from a $d$-dimensional theory to a $(d+1)$-dimensional topological bulk only changes the bosonic dimension. In the supersymmetric case, instead, one must also match the fermionic structure of bulk and boundary in a way compatible with the preserved supercharges. 

Although the formalism is completely general, in this paper we illustrate it in two simple two-dimensional examples. The first one is the $\mathcal{N}=(1,1)$ super-compact boson, which provides a convenient laboratory to analyze global and spacetime symmetries in superspace. The second one is its chiral counterpart, where chirality leads to a different anomaly structure. These two examples are simple enough to keep the discussion transparent, while already exhibiting the main ingredients of the general construction.
In this paper we limit ourselves mainly to consider global symmetries, but the construction can be naturally generalized to include the supersymmetry algebra itself \cite{us}. 

\paragraph*{Outline of the paper.}
The rest of the paper is organized as follows. In section 2 we develop the general construction of SymTFT in superspace and introduce the notion of SuSymTFT for a theory on $\mathcal{SM}^{(d|m)}$ in terms of a bulk topological theory on $\mathcal{SM}^{(d+1|n)}$. In section 3 we apply the formalism to the $\mathcal{N}=(1,1)$ super-compact boson, while in section 4 we discuss the chiral case. We conclude with a discussion of open questions and possible future directions. The paper is supported by four appendices, where we first recall notations and conventions for the two-dimensional ${\cal N}=(1,1)$ supermanifold together with main information about Picture Changing Operators in supergeometry, we then provide details for the evaluation of the mixed anomaly of section 3 and finally we discuss the supersymmetry invariance of the SuSymTFT for the compact boson.

\section{\texorpdfstring{S\lowercase{ym}TFT in superspace}{SymTFT in superspace}}

The symmetry sector of a QFT defined on a $d$-dimensional smooth manifold $\mathcal{M}^{(d)}$ can be encoded in a topological theory in one dimension higher: its Symmetry Topological Field Theory (S\lowercase{ym}TFT). If the  physical theory on $\mathcal{M}^{(d)}$ has a symmetry group $G$ (this may be internal or even spacetime), then its SymTFT is a theory of flat connection for the $G$ group on $\mathcal{M}^{(d+1)}$. If $G$ is finitely generated, then such a topological theory is a BF theory, possibly completed by Chern-Simons terms to capture the full set of  global anomalies.  
The topological BF theory is described by the following general action 
\begin{equation}
\label{gcA}
S_{\rm BF} = \int_{\mathcal{M}^{(d+1)}} b^{(d-p)} \wedge d a^{(p)}
\end{equation}
where $b^{(d-p)}$ and $a^{(p)}$ are continuous abelian $(d-p)$-form and $p$-form, respectively. 

$S_{\rm BF}$ is invariant under gauge transformations 
\begin{equation}
\label{gcB}
\delta a^{(p)} = d \lambda^{(p-1)}\,, ~~~~~~~~
\delta b^{(d-p)} = d \rho^{(d-p-1)}
\end{equation}
If $p>1$ and/or $(d-p-1)>0$, there is a residual gauge symmetry acting on the gauge parameters as well. In addition, the BF action enjoys a $(d-p-1)$-form symmetry with conserved current $J^{(p+1)} = d a^{(p)}$ and a $(p-1)$-form symmetry generated by $\tilde{J}^{(d-p+1)} = db^{(d-p)}$. 

In order to realize the ''sandwich construction'' encoded in a symTFT,  the $(d+1)$-dimensional manifold has the structure $\mathcal{M}^{(d+1)} = \mathcal{M}^{(d)} \times {\cal I}$, where the two extremes of the ${\cal I}$ interval define the symmetry and the physical spaces of the original $d$-dimensional theory. The construction has to be completed with suitable boundary conditions on $a^{(p)}$ and $b^{(d-p)}$ at the physical and topological symmetry boundaries. According to the choice of Neumann or Dirichlet boundary conditions at the symmetry boundary, topological Wilson-like operators of the BF theory give rise to the symmetry generators or non-topological operators charged under that symmetry. 
In the present example, the $d$-dimensional physical theory exhibits a global $(p-1)$-form symmetry generated by a conserved 
current $J^{(d-p)}$  that couples to $a^{(p)}$, and a dual $(d-p-1)$-form symmetry whose conserved current 
$\tilde{J}^{(p)}$ is gauged by $b^{(d-p)}$. 

In the $p=1$ case, this construction admits a straightforward non-abelian generalization\footnote{When $p>1$ one may still realize a non-abelian connection via a higher-group structure \cite{Cremonini:2024mmf}.}. The non-abelian BF theory is simply obtained by covariantizing the action in \eqref{gcA} as
\begin{equation}
\label{gcAna}
S_{\rm BF} = {\rm Tr}\, \int_{\mathcal{M}^{(d+1)}}  b^{(d-1)} \wedge F^{(2)} \; ,  
\qquad {\rm with} \quad F^{(2)} = d a^{(1)} + a^{(1)} \wedge a^{(1)} \equiv {\cal D} a^{(1)}
\end{equation}
where $b^{(d-1)}$ is valued in the adjoint representation of $G$. 
The covariant differential ${\cal D}$ makes this expression invariant under non-abelian gauge transformations 
\begin{equation}
\label{gcBna}
\delta a^{(1)} = {\cal D} \lambda^{(0)}\,, ~~~~~~~~
\delta b^{(d-1)} = {\cal D} \rho^{(d-2)}
\end{equation}
Complemented with suitable boundaries conditions at the two boundaries of ${\mathcal{M}^{(d+1)}}$, the action in \eqref{gcAna} provides the symTFT for a $d$-dimensional physical theory with a non-abelian continuous 0-form symmetry.

\subsection{BF theory on supermanifolds}\label{sec:BF}

We now move to the main problem of constructing the symTFT for  supersymmetric QFTs. This requires first generalizing the BF action to a supersymmetric one. 

The natural geometric setting for studying supersymmetric theories in $d$-dimensions is a supermanifold $\mathcal{SM}^{(d|m)}$, where the fermionic dimension $m$ is given by the number of spinor components, that is the number of supersymmetries times the adapted dimensions of a spinor in the given bosonic dimension $d$ and with a suitable signature\footnote{For instance, a four-dimensional   $\mathcal{N}=1$ theory can be formulated in $\mathcal{SM}^{(4|4)}$, 
with four bosonic dimensions and four fermionic dimensions. Analogously, a four-dimensional  $\mathcal{N}=2$ theory requires  $\mathcal{SM}^{(4|8)}$, and so on.}. We will consider a genus zero, real, smooth, oriented supermanifold, with a riemannian metric and a spin (more  generally spin$^c$) structure\footnote{Topological notions for supermanifolds are by definition the ones for the bosonic submanifold \cite{Witten:2012bg}.}.   Locally, it can be parametrized by a set of $d$ real even coordinates $x^a$, and $m$ odd ones $\theta^\alpha$. Its geometry is described by supervielbeins $(V^a, \psi^\alpha)$, in terms of which one can define the super-differential $d = V^a \partial_a + \psi^\alpha D_\alpha$, with $D_\alpha$ being supersymmetry covariant derivatives. 

On a smooth supermanifold one can define pseudo-forms $\omega^{(p|q)}$, $0\leq q \leq m$, which are featured by a form number $p$ and a picture number $q$. While $p$ is the analogue of the ordinary form degree, $q$ can be changed by multiplying the pseudo-form by a {\em Picture Changing Operator} 
(PCO). We give details in Appendix \ref{sec:PCO}. 

For top forms $\omega^{(p|q=m)}$ a consistent  integration theory has been developed \cite{BL, Witten:2012bg}.\footnote{For a short summary, see for example appendix A of \cite{Grassi:2025tfp}.} These ``integral forms'' are dual to superforms $\omega^{(d-p|q=0)} \equiv \star \omega^{(p|q=m)}$, where in the presence of a metric, the Hodge dual defines isomorphisms. 

In order to write a BF action in the $\mathcal{SM}^{(d+1|m)}$ supermanifold, it is sufficient to promote ordinary forms in the action \eqref{gcA} to pseudoforms, such that their product saturates both the even and the odd dimensions. Taking into account that the differential has picture zero, the most general expression for a super-BF action is then
\begin{equation}
\label{gcC}
S_{\rm sBF} = \int_{\mathcal{SM}^{(d+1|m)}}  b^{(d-p|m-q)} \wedge d a^{(p|q)}
\end{equation}

In the absence of boundaries the equations of motion imply $ d a^{(p|q)}=d b^{(d-p|m-q)}=0$. Moreover, the action is invariant under abelian supergauge transformations of the form
\begin{equation}
 \delta a^{(p|q)} = d\lambda^{(p-1|q)} \, \qquad 
 \delta b^{(d-p|m-q)} = d \rho^{(d-p-1|m-q)}
\end{equation}

Special cases correspond to $q=0$ and $q=m$, where one of the two gauge forms is an integral form and the other one is a superform. \\
We note that, making use of suitable PCOs, we can construct a well-defined action also for pairs of pseudo-forms whose picture number does not saturate the odd dimensions of the supermanifold. In fact, the action
\begin{equation}
\label{gcD}
S_{\rm sBF} = \int_{\mathcal{SM}^{(d+1|m)}}  \!\!\!\!\! b^{(d-p|r)} \wedge d a^{(p|q)} \wedge \mathbb{Y}^{(0|m-q-r)}
\end{equation}
is perfectly well-defined and coincides with \eqref{gcC} under the identification $b^{(d-p|r)} \wedge \mathbb{Y}^{(0|m-q-r)} \equiv b^{(d-p|m-q)}$ (up to a possible sign).
The role of the PCO is to project the integral on a $(d|q+r)$-submanifold. 

In the absence of boundaries, the action in \eqref{gcD} is independent of its embedding. In fact, changing the embedding accounts for a shift of $\mathbb{Y}^{(0|m-q-r)}$ by a $d$-exact term. However, since the integrand $b^{(d-p|r)} \wedge d a^{(p|q)}$ is closed\footnote{As a consequence of the fact that the total integrand is a top form - then closed - and the PCO is closed as well.} this shift is integrated to zero. 

In order to make the previous construction more concrete, we study in details the action for the simple case of ${\cal N} = 1$ super-BF theory in three dimensions ({\em i.e.} $d=2$, $m=2$). 
To shorten the notation, in \eqref{gcC} we set $b^{(1|2)} \equiv B$ and $a^{(1|0)} \equiv \widetilde B $. We work out the action in components by looking at the component expansion of $\widetilde B $, its superfield strength 
$\widetilde{F} = d \widetilde{B}$ 
and the integral form $B$, separately. 
The $\widetilde{B}$ superform and its field strength $\widetilde{F}$ can be expanded on the basis of the $(V^a, \psi^\alpha)$ super-dreibeins as follows
\begin{equation}
\label{Ftilde} 
\widetilde B=\widetilde B_aV^a+\widetilde B_\alpha \psi^\alpha ~~~~~~~~~\widetilde{F} =  \widetilde{F}_{ab} V^a V^b + (\widetilde{W}\gamma_a \psi)V^a 
\end{equation}
 where, in order to select only the vector multiplet in its expansion, we  imposed some constraints (a.k.a. {\it conventional constraints}) on $\widetilde F$. 
In the above expansion, $\widetilde{F}_{ab}$ is the superfield strength in superspace, whereas the $\widetilde{W}$ superfield  has the gaugino as its lowest component. Bianchi identities together with  conventional constraints imply $D^\alpha \widetilde{W}_\alpha =0$ and $\widetilde{F}_{ab} = D \gamma_{ab} \widetilde{W}$. 

In order to write the expansion for $B$ we first introduce the (1|0)-superform ${\mathcal B}$ and its superform field strength ${\mathcal F} = d{\mathcal B}$, whose component expansion is given by
\begin{equation}
\label{F} 
 {\mathcal B}= B_aV^a+ B_\alpha \psi^\alpha ~~~~~~~~~{\mathcal F}=  F_{ab} V^a V^b + (W\gamma_a \psi)V^a 
\end{equation} 
where conventional constraints have been imposed, in analogy with the $\widetilde{F}$ superform.
Now, upon a suitable gauge fixing, the $(1|2)$-integral form $B$ can be written in terms of these pseudoforms as 
\begin{equation}
    B^{(1|2)}=\left({\mathcal B}+\frac{1}{3!}V^a V^b W\gamma_{ab}\iota\right)\wedge \mathbb{Y}^{(0|2)}
\end{equation}
for an arbitrary PCO $\mathbb{Y}^{(0|2)}$. The presence of the $W$-term is crucial for ensuring the closure of $B$. 

Inserting the previous expansions in equation \eqref{gcC}, the action of the super-BF model can be cast in the following form 
(see \cite{Grassi:2016apf} for details) 
\begin{equation}
\begin{split}
    \label{hoI}
    S_{\rm sBF}
    &=  \int_{\mathcal{SM}^{(3|2)}}B\wedge \widetilde F=\int_{\mathcal{SM}^{(3|2)}} \Big( {\mathcal B} \wedge\widetilde{F} + \epsilon^{\alpha \beta}W_{\alpha} \widetilde{W}_\beta V_3\Big) \wedge \mathbb{Y}^{(0|2)}
\end{split}
\end{equation}
This is the super-BF theory written in the supergeometry language.  
The integrand factorizes into a closed (3|0)-superform Lagrangian and a generic (0|2) PCO. 

Given the closure of the Lagrangian,  
in \eqref{hoI} we can choose different $\mathbb{Y}^{(0|2)}$ without modifying the action. For example, if we choose 
$\mathbb{Y}^{(0|2)} = \theta^2 \delta^2(\psi)$, the super-BF action reduces to its component form 
\begin{equation}
\begin{split}
    \label{hoIA}
    S_{\rm sBF}
    &=  \int d^3x \, \Big( \epsilon^{abc} b_a \partial_b \widetilde{b}_c + \lambda^B_\alpha \epsilon^{\alpha \beta} \lambda^{\tilde B}_\beta\Big) 
\end{split}
\end{equation}
where $b_a= B_a|_{\theta=0}$, and $\tilde{b}_a=\widetilde B_a|_{\theta=0}$ are the component gauge fields of the superforms $B$ and $\widetilde{B}$ respectively, while  $\lambda^B=W|_{\theta=0}$ and $\lambda^{\tilde B}=\widetilde{W}|_{\theta=0}$ are their supersymmetric partners. 

\vskip 5pt

We conclude this section by observing that, as in the bosonic case, for $p=1$ we can generalize the abelian BF theory on supermanifolds to be non-abelian.

Taking inspiration from \eqref{gcAna}, we write the most general non-abelian BF theory in supermanifold as
\begin{equation}
S_{\rm sBF} = {\rm Tr}\,\int_{\mathcal{SM}^{(d+1|m)}}  \hspace{-0.7cm} b^{(d-1|m)} \wedge F^{(2|0)} \qquad
 {\rm with}
\quad  F^{(2|0)} = da^{(1|0)} + a^{(1|0)} \wedge a^{(1|0)} 
\end{equation} 
where again $b_a = B_a^{(1|0)}|$ takes values in the adjoint representation of $G$.
This action is invariant under the following gauge transformations
\begin{equation}
\label{gcBnaB}
\begin{split}
\delta a^{(1|0)} &={\cal D} \lambda^{(0|0)}\,, \quad
\delta b^{(d-1|m)} = {\cal D} \rho^{(d-2|m)}
\end{split}
\end{equation}
where ${\cal D} =  d +  
[a^{(1|0)}, \cdot \, ]$.

We note that in the non-abelian case, the non-linear dependence on the gauge superconnection forces the field strength $F^{(2|0)}$ to be necessarily a superform ($q=0$). However, we can always use a PCO to change its picture. Therefore, defining a generalized field strength 
\begin{equation}
 F^{(2|q)} = F^{(2|0)} \wedge \mathbb{Y}^{(0|q)}  \end{equation}
a more general non-abelian BF action takes the form
\begin{equation}
S_{\rm sBF} = {\rm Tr}\,\int_{\mathcal{SM}^{(d+1|m)}}  \hspace{-0.7cm} b^{(d-1|m-q)} \wedge F^{(2|q)} 
\end{equation}

\subsection{SuSymTFT}
\label{sect:SuSymTFT}

Having introduced the topological BF theory on supermanfolds, we can use it to construct the Supersymmetric version of the SymTFT (SuSymTFT) for supersymmetric theories defined in superspace. This formulation has the advantage of exhibiting manifest supersymmetry, thus conveniently packaging the full set of bosonic and fermionic anomalies. 

Generalizing the construction reviewed above for bosonic theories, in the supersymmetric setup the SuSymTFT for a theory living in a supermanifold $\mathcal{SM}^{(d|m)}$  is a topological field theory on the supermanifold $\mathcal{SM}^{(d+1|n)}$ such that $\partial \mathcal{SM}^{(d+1|n)} = \mathcal{SM}^{(d|m)}$. More precisely, the sandwich construction requires the SuSymTFT to have two boundaries, still dubbed \textit{physical} and \textit{topological}, the former being the one where the physical theory lives and the second one being a topological boundary condition for the bulk fields.  Boundary conditions of the bulk fields at the boundaries guarantee gauge invariance of the bulk theory, at the same time topological boundary conditions specify which absolute theory lives on the physical boundary after interval compactification.

So far, this looks like a straightforward generalization of the bosonic construction. However, in the supersymmetric case special attention has to be devoted to the relation between the odd dimensions of the bulk and the boundary supermanifolds. In fact, since boundaries (partially) break supersymmetry, in general the dimension of the fermionic support of the  $\mathcal{SM}^{(d|m)}$ and $\mathcal{SM}^{(d+1|n)}$ supermanifolds are not the same. Typically, a co-dimension one boundary is 1/2-BPS, preserving half of the supersymmetry charges. In fact, given the location of the boundary at $x^\perp=0$, preserved Killing spinors are the ones which satisfy  $\delta_\epsilon x^\perp=\epsilon\gamma^\perp\theta= 0$. The maximal solution is  obtained by splitting  $\epsilon\gamma^\perp\theta = \epsilon_- \theta_+ - \epsilon_+ \theta_-$ with $\theta_\pm=P_\pm\theta, \epsilon_\pm=P_\pm\epsilon$, $P\pm=\frac{1}{2}(1\pm\gamma^\perp)$, and taking $\epsilon_- =  \theta_- =0$ or $\epsilon_+ =  \theta_+ =0$ or possibly mixed conditions. Therefore, the correct boundary to consider in the SuSymTFT construction can be sketchy defined as 
\begin{equation}
\mathcal{SM}^{(d|m)}  \equiv \partial(\mathcal{SM}^ {(d+1|2m)}) 
=\{x^\perp=0, \theta_+ = 0 \; {\rm or} \; \theta_-=0, {\rm or \, mixed \, conds}  \}
\end{equation} 
where half of the supersymmetries present in $(d+1)$ dimensions survive.

At the level of the action, since in components the supersymmetry  variation of the bulk lagrangian is always a spacetime derivative, half of the supersymmetries can be preserved by imposing suitable boundary conditions on the fields and their derivatives or adding a suitable boundary term to the bulk action, whose variation cancels the residual term. The second approach is the “F+A” mechanism (or “susy without BC”) introduced in \cite{Belyaev:2008xk} to preserve the right amount of supersymmetry in theories with boundaries. However, in the present context it is more natural to impose  boundary conditions on the fields, as we are going to discuss in the explicit examples studied in the next sections.

\section{Scalar Supermultiplet }
As a first simple example, we consider the free theory for a  $\mathcal{N}=(1,1)$ super-compact boson living in a flat supermanifold 
$\mathcal{SM}^{(2|2)}$. We collect notations and conventions for $\mathcal{SM}^{(2|2)}$ in appendix \ref{appA}. 

In the non-supersymmetric case, the construction of the SymTFT for the non-invertible T-duality symmetry of the compact boson has been considered in \cite{Argurio:2024ewp}.

\subsection{The super-compact boson}
In the two-dimensional $\mathcal{N}=(1,1)$ superspace, we consider a real $(0|0)$-superform $\Phi$ whose $\theta$-expansion is given by 
\begin{equation}
\label{eq:exp}
    \Phi \equiv \Phi(z,\bar{z}, \theta, \bar\theta) = \phi(z,\bar{z}) + \theta \lambda(z,\bar{z}) + \overline{\theta \lambda}(z,\bar{z}) + \theta \bar\theta F(z,\bar{z})
\end{equation}

In order to implement compactness, we restrict $\Phi$  to be defined  as a section valued in a target supermanifold  $\mathcal{S}S^{1}$ which possesses one compact bosonic direction
$$ \Phi: \mathcal{SM}^{(2|2)} \to \mathcal{S}S^1$$
The section of $\mathcal{S}S^{1}$ of radius $R$ is defined in such a way that the lowest component of the superfield $\Phi\eval_{\theta = 0} = \phi $ is a map from $ \mathcal{SM}^{(2|2)}$ to the circle $S^1$, periodically identified as 
$$\phi(z,\bar{z}) = \phi(z,\bar{z}) +2 \pi R$$ 
The higher components of $\Phi$ are not periodically identified, being maps to the (super)tangent bundle of $S^1$.

The physical degrees of freedom - one compact scalar plus two Majorana-Weyl fermions - are governed by the following action \cite{Castellani:2015ata}
\begin{equation}
    \label{hoA}
    S= \frac{1}{4\pi}\int_{\mathcal{SM}^{(2|2)}} d\Phi \wedge \star d\Phi
\end{equation}
where  $\star$ is the super-Hodge dual \cite{Castellani:2015dis}. Defining the extra spinorial superfield 
\begin{equation}
W_\alpha \equiv  (W, \overline W) = (D \Phi, \overline D\Phi)
\end{equation}
from the explicit expression \eqref{eq:d} for the differential we find (see appendix \ref{appA})
\begin{equation}
\begin{split}
    \label{hoB}
    d\Phi &= V^a \partial_a \Phi + \psi^\alpha W_\alpha \\
    \star d\Phi &= 
    \epsilon^{ab}  \, V_b \,  \delta^2(\psi) \, \partial_a \Phi
    + V \overline{V} \, \epsilon^{\alpha\beta}\iota_\alpha\delta^2(\psi)  W_\beta 
\end{split}
\end{equation}
where $V^a \equiv (V, \overline V)$ and $\psi^\alpha \equiv (\psi, \overline \psi)$ are the super-vielbeins, and $\iota_\alpha$ is the contraction along the $D_\alpha \equiv (D, \overline{D})$ vector field. 

Since the action in \eqref{hoA} is the integral of a top form, it 
is manifestly supersymmetric as long as the $\mathcal{SM}^{(2|2)}$ supermanifold has no boundary. In fact, taking into account that its supersymmetry variation is defined as  $\delta_\epsilon S \equiv \int [{\cal L}_\epsilon ,d\Phi \wedge \star d\Phi]$ where ${\cal L}_\epsilon = [d, \iota_\epsilon]$ is the Lie derivative in the $\epsilon$ direction, we are left with a $d$-exact term that integrates to zero on a trivial boundary.

The equation of motion from this action is $d  \star d\Phi=0$. Using expansions \eqref{hoB} it can be rewritten as
\begin{equation}
    \label{hoC}
     V \overline{V} \delta^2(\psi) (1 + \epsilon^{\gamma\delta} D_\gamma D_\delta) (\epsilon^{\alpha\beta} D_\alpha D_\beta \Phi ) = 0  
\end{equation}
which implies  
$D\bar{D} \Phi = \bar{D} D \Phi =0$. 
In fact, it is easy to see that on-shell the action \eqref{hoA} reduces to the ordinary superspace action \cite{Gates:1983nr} 
\begin{equation}
    S = \frac{1}{4\pi} \int d^2z d^2\theta \; D \Phi \overline{D} \Phi 
\end{equation}
This action defines a superconformal field theory at the quantum level. 

\subsubsection{Global symmetries}

The system possesses two (on-shell) closed quantities, a $(1|0)$-superform $J = d \Phi$ and its dual $(1|2)$-integral form $\widetilde{J} = \star d \Phi$, which  lead to two  conserved $U(1)$ currents. They correspond to two $U(1)$ symmetries, one $(0|2)$-integral form and  one $(0|0)$-superform symmetries, respectively. In particular,  the $(0|2)$-integral form symmetry corresponds to constant shifts, $\Phi \to \Phi + \omega$, where $\omega$ is a constant superfield. 
The  topological charges implementing the action of these symmetries on local operators are 
\begin{equation}\begin{split}
    \label{hoCA}
    U_g(\Gamma^{(1|0)}) &=\exp( \frac{ig}{2\pi R} \int_{\Gamma^{(1|0)}} \!\! \! d \Phi) =\exp(\frac{ig}{2\pi R} \int_{\mathcal{SM}^{(2|2)}}\!\! \!  d \Phi \wedge \mathbb{Y}^{(1|2)}_{\Gamma^{(1|0)}} )
     \\
    V_h(\Sigma^{(1|2)}) &= \exp(i h R \int_{\Sigma^{(1|2)}} \!\! \! \star d \Phi ) =\exp(i h R\int_{\mathcal{SM}^{(2|2)}}\!\! \!\star d \Phi \wedge \mathbb{Y}^{(1|0)}_{\Sigma^{(1|2)}})  
\end{split}
\end{equation}
where $g,h$ are grassmann even constant parameters satisfying $g,h \sim g,h + 2\pi$. Moreover, $\mathbb{Y}^{(1|2)}_{\Gamma^{(1|0)}}$, $\mathbb{Y}^{(1|0)}_{\Sigma^{(1|2)}}$ are two PCOs which localize the integrals on the corresponding hypersurfaces.
Topological invariance of these charges easily follows from the closure of the currents, together with the crucial property that under smooth deformations of the hypersurfaces the PCOs transform by a $d$-exact term.

The two operators in \eqref{hoCA} are associated respectively to the momentum and winding charges of the supersymmetric vertex operator $:\!\exp(i n \Phi)\!\!:$ of the compact boson CFT.

Because of the compactness of the target space, the holonomies are quantized, the action of the charges on the Hilbert space has a non trivial kernel, and it is generated by the 2-torus
$$ 
(\mathbb{R}/2\pi \mathbb{Z})^{(0|0)}\times (\mathbb{R}/2\pi \mathbb{Z})^{(0|2)}\simeq U(1)^{(0|0)}\times U(1)^{(0|2)}
$$
exactly as in the bosonic case. 

\vskip 5pt
It is interesting to observe that in the presence of supersymmetry $J, \widetilde{J}$ do not exhaust the spectrum of global $U(1)$ currents. In fact, they belong to two complete supermultiplets, whose components are still superform and integral form currents, respectively. For instance, if we apply a supersymmetry transformation to $J$ we obtain a closed spinorial current ${\mathcal J}_\alpha = ({\mathcal J}, \overline{\mathcal J})$, with
\begin{equation}
\label{eq:susytransformedcurrents}
\epsilon  {\mathcal J} \equiv \delta_\epsilon J =
{\mathcal L}_\epsilon J  
 \; , \qquad \quad 
\bar\epsilon  \overline{\mathcal{J}} \equiv \delta_{\bar\epsilon} J =
{\mathcal L}_{\bar\epsilon} J 
\end{equation}
where ${\mathcal L}_{\epsilon, \bar\epsilon}$ are the Lie derivatives in the $\epsilon, \bar\epsilon$ directions, respectively. 
The closure of this new current easily follows from $[d,{\mathcal L}_{\epsilon, \overline{\epsilon}} ] = 0$. 

Analogously, applying a supersymmetry transformation to $\widetilde{J}$ we generate another spinorial closed current
$\widetilde{\mathcal J}_\alpha = (\widetilde{\mathcal J}, \overline{\widetilde{\mathcal J}})$, where $\widetilde{\mathcal J} = \star {\mathcal J}$ and $\overline{\widetilde{\mathcal J}} = \star \overline{\mathcal J}$.
A simple calculation shows that 
\begin{equation}
    \label{compA}
    \mathbb{Z} \overline{\mathbb{Z}}[\widetilde{\mathcal J}] = {\mathcal J} \,, ~~~~~~~~~~~
    \mathbb{Z} \overline{\mathbb{Z}}[ \overline{\widetilde{\mathcal J}}] = \overline{\mathcal J} 
\end{equation}

Additional supercurrents can be constructed as $\kappa$-symmetry superpartners, by acting on $J$ with the Lie derivative with respect to the vector $\kappa = \kappa^\alpha D_\alpha$.  We obtain another spinor of currents $(\mathcal{J}_\kappa, \overline{\mathcal{J}}_\kappa)$, where 
\begin{equation}
\label{Kcurrents}
\mathcal{J}_\kappa = d W \, , \qquad \overline{\mathcal{J}}_\kappa = d \overline W
\end{equation}
They are trivially closed, consistently with the fact that in the present geometric framework $\kappa$-symmetry is a super-diffeomorphism. 

The $\kappa$-symmetry transformed currents generate constant shifts of the fermions, $W \to W + c$, $\overline{W} \to \overline{W} + \overline{c}$, where $(c, \overline{c})$ are constant superfields. 
It is important to note that the fermion shift is not independent of the boson one, $\Phi \to \Phi + \omega$. In fact, since $(W, \overline{W}) = (D\Phi, \overline{D} \Phi)$, we trivially have $(c, \overline{c}) = (D\omega, \overline{D} \overline\omega)$. In other words, the parameters  are related by supersymmetry transformations as 
\begin{equation}
    \label{shAAA}
\delta_\epsilon \omega = \epsilon c \, , \quad \delta_\epsilon c =0 \, ;  \quad \quad \delta_{\bar\epsilon} \overline{\omega} = \bar\epsilon \overline{c} \, , \quad \delta_{\bar\epsilon} \overline{c} =0
\end{equation}

\subsubsection{Spacetime symmetries}

We now discuss the spacetime symmetries of the super-compact boson. Although in two dimensions the $\mathcal{N} = (1,1)$ superconformal group is enhanced to the infinite dimensional super-Virasoro algebra, here we focus only on super-translations.
Since supersymmetry and spacetime symmetries are naturally identified as  $(0|0)$-superform symmetries \cite{Grassi:2025tfp}, they should correspond to $(1|2)$-form closed currents. Indeed, a supersymmetry spinorial $(1|2)$-current can be constructed, whose components are given by 
\begin{equation}
    \label{sussaA}
    \mathcal{S}^{(1|2)} =  
    W \partial \Phi \, V \delta(\psi) \wedge \overline{\mathbb{Y}}^{(0|1)} \, , \qquad \quad 
    \overline{\mathcal{S}}^{(1|2)} =
    \overline W \overline \partial \Phi \, \overline V \delta(\overline \psi)\wedge  {\mathbb{Y}}^{(0|1)}
    \end{equation}
upon a suitable choice of the PCOs. These two components  can be checked to be  closed using the on-shell free equations 
$\partial\bar\partial \Phi =0, D \overline{W} =0$ and $\overline{D} W=0$. 

Supersymmetry currents with lower picture can be obtained by applying the  $\mathbb{Z}$, $\overline{\mathbb{Z}}$ PCOs defined in appendix \ref{sec:PCO}. Precisely, we can consider the alternative but equivalent currents\footnote{We can choose ${\mathbb{Y}}^{(0|1)}= V \iota \delta(\psi)$ and use $\mathbb{Z}[V \delta(\psi)] = - \psi$,  
$\mathbb{Z}[V \delta'(\psi)] = 1$, and analogous relations for the anti-holomorphic sector.} 
\begin{equation}\label{S11}
    \mathcal{S}^{(1|1)} \equiv \overline{\mathbb Z} \mathcal{S}^{(1|2)} = W \partial \Phi \, V \delta(\psi) \, , \qquad \quad
    \overline{\mathcal{S}}^{(1|1)} \equiv {\mathbb Z} \overline{\mathcal{S}}^{(1|2)} = \overline W \overline \partial \Phi \, \overline V \delta(\overline \psi)
\end{equation}
or, applying both PCOs, a third pair of closed currents (we insert a minus sign for convenience)
\begin{equation}
\begin{split}
    \label{sussaB}
&\hspace{-0.7cm} \mathcal{S}^{(1|0)} \equiv 
     -{\mathbb Z} \overline{\mathbb Z} \mathcal{S}^{(1|2)} 
    \! = 
     W \partial \Phi \, \psi + ( (\partial \, \Phi)^2 + W\partial W) V
      \\
& \hspace{-0.7cm}\overline{\mathcal{S}}^{(1|0)}  
\equiv  -{\mathbb Z} \overline{\mathbb Z} \overline{\mathcal{S}}^{(1|2)}
\! = \overline{W} \overline\partial \Phi \, \overline\psi +  ( (\bar\partial \Phi)^2 + \overline W\bar \partial \overline W) \overline{V} 
     \end{split}
     \end{equation}
Here, the $\psi, \overline{\psi}$ and $V, \overline{V}$ components are easily recognized to be the holomorphic and antiholomorphic parts of the supersymmetry current and the energy momentum tensor respectively,  written in superfields. 
If we further  expand $W \partial \Phi$ and $\overline{W} \overline\partial \Phi$ in components by the use of \eqref{eq:exp}, 
\begin{equation}
\begin{split}
    \label{muD}
  &\hspace{-0.3cm} W \partial \Phi = \Big[ \lambda \partial \phi - 
    \theta \Big( (\partial \phi)^2 + \lambda \partial \lambda\Big) \Big]  = j_S - \theta \, T \\
   &\hspace{-0.3cm}\overline{W} \overline\partial \Phi = \Big[ \overline\lambda \overline\partial \phi - 
    \overline\theta \Big( (\overline\partial \phi)^2 + \overline\lambda \overline\partial \overline\lambda\Big) \Big] = \overline{j}_S - \overline\theta \, \overline{T}
\end{split}
\end{equation}
we recognize the ordinary holomorphic component of supersymmetry current, $j_S = \lambda \partial \phi$, and the holomorphic energy-momentum tensor $T = (\partial \phi)^2 + \lambda \partial \lambda $, plus their antiholomorphic counterparts. 

In order to determine the  topological generators of super-translations, it is convenient to use the $(1|2)$-picture of the $S$ current, see eq.  \eqref{sussaA}, and first make an explicit choice of the two PCOs. Here we choose ${\mathbb{Y}}^{(0|1)}= V \delta'(\psi), \overline{\mathbb{Y}}^{(0|1)} = \overline V  \delta'(\bar \psi)$.
We then couple the currents to two super-parameters  with constant components, $A=a+\epsilon\theta$  and $\bar{A}=\bar a+\overline{\epsilon\theta}$, and integrate on the (1|2)-surfaces $\Sigma:=\{ \bar z=0\}$ and $\overline{\Sigma}:=\{  z=0\}$, respectively. The projection on these  two surfaces is realized by the two PCOs $\overline{\mathbb{Y}}^{(1|0)} = d\bar z\delta(\bar z) $ and ${\mathbb{Y}}^{(1|0)} = dz\delta(z)$.  Performing the supermanifold and the residual Berezin integrals explicitly, we eventually obtain the infinitesimal transformation operators 
\begin{equation}
\begin{split}
\label{susycharges}
\delta_A &= \!\! \int_\Sigma  A \mathcal{S}^{(1|2)}= 
a \!\!\int dz \, T(z) + \epsilon \!\! \int dz \, j_S(z) = a P + \epsilon Q
\\
\delta_{\bar{A}} &= \! \! \int_{\overline\Sigma} \overline{A} \overline{\mathcal{S}}^{(1|2)}  
= \overline{a} \!\! \int d\bar{z} \, \overline{T}(\bar{z}) + \overline\epsilon \!\! \int d \bar{z} \, \overline{j}_S(\bar{z}) = \bar{a} \overline{P} + \overline\epsilon \overline{Q}
\end{split}
\end{equation}
The $a, \overline{a}$ charges are nothing but the usual translation parameters multiplying the translation operators in direction $z$ and $\bar{z}$, whereas $\epsilon, \overline{\epsilon}$ are two Killing spinors multiplying the supersymmmetry charges in the two chiral sectors.

Under the action of the supersymmetry operator $Q$ the components of the current multiplet transform as 
\begin{equation}
    \label{muE}
    \delta_\epsilon j_S = -\epsilon \, T\,, ~~~~~
    \delta_\epsilon T =  \epsilon \, \partial j_S
\end{equation}
Analogous identities hold in the antichiral sector. 

The construction of all Virasoro generators follows a similar pattern and will be reported elsewhere \cite{us}.

\subsubsection{Super-Chern-Weil symmetries}

As discussed in \cite{Grassi:2025tfp}, we expect the emergence of additional bosonic and fermionic higher form symmetries. For example, given the currents $J= d\Phi$ and  $\mathcal{J}_{\kappa} = dW, \overline{\mathcal{J}_{\kappa}} = d \overline{W}$  
we can construct the following higher conserved currents 
\begin{eqnarray}
    \label{higA}
    J^{(2|0)}_{bf} &=& d\Phi \wedge d W\,,  ~~~~
    J^{(2|0)}_{b\bar f} = d\Phi \wedge d \overline{W}, \nonumber \\    
    J^{(2|0)}_{ff} &=& dW \wedge d W\,, ~~~~J^{(2|0)}_{f\bar f} = dW \wedge d \overline{W}\,, ~~~~
J^{(2|0)}_{\bar f\bar f} = d\overline{W} \wedge d\overline{W}
\end{eqnarray}
In contrast with the previous construction, these currents are not integral forms on the supermanifold, as they do not saturate the odd directions. Therefore, the definition of the corresponding topological charges requires performing a suitable projection in the fermionic directions. On the other hand, since these currents are spacetime filling,  they are expected to give rise to some super-version of $(-1)$-form symmetries.  

In addition, we can construct other $(2|0)$-form currents by exploiting the two Chevalley-Eilenberg cohomologies of $\mathcal N=(1,1)$ superspace (see \cite{Grassi:2025tfp, Cremonini:2022cdm}), namely $\omega^{(1|0)} = \psi$ 
and  $\overline\omega^{(1|0)} = \overline\psi$.  New higher-form currents are
\begin{eqnarray}
    \label{higB}
    &&J_{\omega b}^{(2|0)} = \psi \wedge d\Phi, ~~~~
    J_{\bar \omega b}^{(2|0)} = \bar\psi \wedge d\Phi,
   \\
    &&
    J_{\omega f}^{(2|0)} = \psi \wedge dW, ~~~~
    J_{\omega \bar f}^{(2|0)} = \psi \wedge d\overline W, ~~~~
    J_{\bar\omega f}^{(2|0)} = \bar\psi \wedge dW, ~~~~
    J_{\bar\omega \bar f}^{(2|0)} = \bar\psi \wedge d\overline W \nonumber
\end{eqnarray}

Following this path, we find several new higher-form currents (even and odd) which might play an interesting role, another example being 
$J^{(4|0)}= \psi \wedge\bar \psi\wedge dW\wedge dW$. However, at the moment their r\^ole in physics is not clear. We leave this issue to future analysis. 

\subsection{Mixed anomaly}

As a preparatory work towards the construction of the susymTFT for the super-compact boson, we investigate the emergence of possible 't Hooft anomalies for the various currents. 

To this end, we adopt a well-established approach. We first  minimally couple the symmetry currents to background gauge fields. Contextually, we use one of the background fields to covariantize the currents with respect to the corresponding symmetry - in this case we choose the shift symmetry - then we require the background to transform non-trivially under gauge transformations in order to ensure gauge invariance of the total action, up to a possible term that depends only on the background fields and the gauge parameters. The appearance of such a left-over signals the 
presence of a 't Hooft anomaly which can be compensated by a suitable inflow action. 

Focusing only on the global and spacetime symmetries, we minimally couple  $J = d\Phi$ and $\widetilde{J} = \star d\Phi$ to background gauge fields $B^{(1|0)} \equiv B$ and $\widetilde{B}^{(1|2)}\equiv \widetilde{B}$ respectively, and the supersymmetry current ${\mathcal S}^{(1|1)}$ in \eqref{S11} to a gravitino background $\chi^{(1|1)} \equiv \chi$. 
Moreover,
in order to make the action  manifestly invariant under the shift symmetry we use the $B$ field to covariantize $J$ and ${\mathcal S}^{(1|1)}$.

The general expansion of the $B$ superform in components reads
\begin{equation}
    \label{newA}
    B= b_V V + \overline{b}_V \overline V + b_\psi \psi + \overline b_\psi \overline \psi
\end{equation}
where $b_V, 
\overline{b}_V, b_\psi,  \overline{b}_\psi$ are ordinary superfields. Imposing conventional constraints on its field strength $F^{(2|0)} = dB$, that is imposing the vanishing of its $\psi^2$ and $\overline\psi^2$ components, it is easy to see that the component superfields of $B$ are not all independent. Precisely, they are related as $b_V = - Db_\psi, \overline{b}_V = - \overline{D} \overline b_\psi$.

The covariantized version of the supersymmetry current can be obtained by exploiting the $B$ components. It reads 
\begin{equation}
\label{covcurrents}
{\mathcal S}_{\rm cov} = (W - b_\psi) (\partial \Phi - b_V) V \delta(\psi) ~~~,~~~  \overline {\mathcal S}_{\rm cov} = (\overline W - \overline{b}_\psi) (\overline \partial \Phi - \overline{b}_V)\overline V \delta(\overline \psi) \end{equation}

We then promote the original action
\eqref{hoA} to
\begin{equation}
    \label{hoD}
    S[B, \widetilde{B}, \chi] = \frac{1}{4\pi} \int_{_{\mathcal{SM}^{(2|2)}}} \Big[ (d\Phi - B) \wedge \, \star (d\Phi- B) + 
    \widetilde{B} \wedge (d\Phi - B) 
+ \chi \left( {\mathcal S}_{\rm cov} + \overline{\mathcal S}_{\rm cov} \right) \Big] 
\end{equation}
Note that the coupling to the background fields does not spoil the manifest supersymmetry of the original action, being $S[B, \widetilde{B}, \chi]$ the integral of a top form on a supermanifold without boundary. 

It is easy to see that the gauged  action is manifestly invariant under local shifts 
\begin{equation}
 \Phi \to \Phi +  \omega \, , \qquad B \to B +  d\omega    
\end{equation}
Moreover, as detailed in appendix \ref{app:mix-anomaly-details}, applying gauge transformations of  gravitational and $\widetilde{B}$ backgrounds of the form
\begin{equation}
    \chi \to \chi + d \Lambda \,  , ~~\qquad 
    \widetilde{B} \to \tilde{B} + d\widetilde{\omega} - \Sigma(\Lambda, B)
\end{equation}
with $\Sigma(\Lambda, B)$ determined in appendix \ref{app:mix-anomaly-details} (see eqs. (\ref{sigma},\ref{sigmacomponents})), the action is invariant up to the following 't Hooft anomaly 
\begin{equation}
     \label{hoE} \delta S[B,\widetilde{B}, \chi] =  
    \frac{1} {4\pi}\int_{\mathcal{SM}^{(2|2)}}   \!\! \! \widetilde\omega \wedge dB 
 \end{equation}
This is the supersymmetric version of the $U(1)^{(0|0)}\times U(1)^{(0|2)}$ mixed anomaly, already present in the bosonic case . It prevents the two global symmetries from being simultaneously gauged. We note that there is no gravitational one-loop anomaly, as expected for a non-chiral theory. Therefore, expression \eqref{hoE} exhausts the spectrum of anomalies of the theory. 


The 't Hooft  anomaly can be canceled by adding to $S[B,\widetilde{B}, \chi]$ the following inflow action 
\begin{equation}
    \label{hoF}
    S_{\rm inflow}[B, \widetilde{C}] =  -\frac{1}{4\pi}\int_{\mathcal{SM}^{(3|4)}}  \widetilde{C}  \wedge d B
\end{equation}
defined in the supermanifold $\mathcal{SM}^{(3|4)}$ with one bosonic dimension higher and twice the number of supercharges, whose boundary is $\mathcal{SM}^{(2|2)}$. $\widetilde{C}$ is a gauge $(1|4)$-superform in the bulk subject to gauge transformations $\widetilde{C} \to \widetilde{C} + d\widetilde{\omega}_B$, with boundary condition $\widetilde{\omega}_B| = \widetilde{\omega}$.

The inflow is exactly the  super-BF action introduced in section \ref{sec:BF}, specialized to the $(3|4)$-dimensional case.

\subsection{SuSymTFT for the super-compact boson}\label{sect:symTFTcompactboson}

In this section we construct the Supersymmetric-SymTFT (SuSymTFT) for the two-dimensional non-chiral $\mathcal{N}=(1,1)$ compact boson. We consider only the global $U(1)^{(0|0)} \times U(1)^{(0|2)}$ symmetries, postponing the construction for spacetime (super)symmetries to a future paper \cite{us}.

In order to define a consistent bulk three-dimensional topological theory coupled to this model, we follow the general prescription proposed in section \ref{sect:SuSymTFT}. 
As already mentioned there, the only non-trivial question concerns the number of supersymmetries that we need to realize in three dimensions. Since the boundary halves the number of supersymmetries, the correct supermanifold where defining the SuSymTFT in this case is  $\mathcal{SM}^{(3|4)}$. In other words, the SuSymTFT is going to be a three-dimensional $\mathcal{N}=2$ super-BF theory.

The continuous global symmetry sector is encoded into a SuSymTFT consisting of a bulk abelian $\mathcal{N}=2$ supersymmetric BF theory, whose action in the   $\mathcal{SM}^{(3|4)}$ supermanifold reads (conventions in three dimensions are given in appendix \ref{appD})
\begin{equation}\begin{split}\label{susycompacglobal}
S_{\rm sBF} &= \frac{1}{4\pi} \int_{\mathcal{SM}^{(3|4)}} \!\!\mathcal{A}_1 \wedge d\mathcal{A}_2 
\end{split} 
\end{equation}
where the fields $\mathcal{A}_I$  are flat $\mathbb{R}$-valued superform connections,
\begin{equation} 
\mathcal{A}_1 \in \Omega^{(1|4)}(\mathcal{SM}^{(3|4)},\mathbb{R})\; , \quad \mathcal{A}_2 \in \Omega^{(1|0)}(\mathcal{SM}^{(3|4)},\mathbb{R})
\nonumber
\end{equation}
subject to the gauge transformations $\delta \mathcal{A}_I = da_I$, for appropriate form degrees. The bulk action needs then be completed with the boundary term $S[B, \widetilde{B},\chi]$ in \eqref{hoA}.\\
The only non-trivial observables of the theory are  $\mathcal{A}_I$ holonomies over non-trivial cycles in $\mathcal{SM}^{(3|4)}$,   measured by the topological operators 
\begin{equation}
  \mathcal{U}_x(\Gamma) = \exp(i x \int_{\Gamma^{(1|4)}}\!\!\!\! \mathcal{A}_1),      \, \qquad
\mathcal{V}_y(\Sigma) = \exp(i y \int_{\Sigma^{(1|0)}} \!\!\!\! \mathcal{A}_2)
\end{equation}
where, for the sake of notation, we dropped the upper-script  $(\cdot|\cdot)$.
 Because of the pairing in \eqref{susycompacglobal}, the operators braid non trivially 
 \begin{equation}\label{nonchibraiding} {\cal U}_x(\Gamma) \, {\cal V}_y(\Sigma)  = e^{4\pi i xy  \,\mathcal{SL}(\Gamma,\Sigma)} \,
 \mathcal V_y(\Sigma)\, \mathcal U_x(\Gamma) \end{equation}
where $\mathcal{SL}(\Gamma, \Sigma)$ is the super-linking number between $\Gamma$ and $\Sigma$, cfr.\ \cite{Grassi:2025tfp}.

To make this super-BF theory the SuSymTFT  of the super-compact boson, we have to fix appropriate physical and topological boundary conditions. 
To this end, we take  $\mathcal{SM}^{(3|4)}$ to be a supermanifold of the form
$$\mathcal{SM}^{(3|4)} = \mathcal{I}^{(1|2)} \times \left(\mathcal{SM}^{(2|2)}_{\rm phys}\cup \overline{\mathcal{SM}}_{\rm  top}^{(2|2)}\right)$$
where $\mathcal{I}$ is a super-interval (a  supermanifold whose bosonic component is an interval) and $\mathcal{SM}^{(2|2)}_{\rm phys} = \mathcal{SM}^{(2|2)}$ is the supermanifold where the super-compact boson is defined. 

At the physical boundary we impose the following boundary conditions for the bulk gauge superforms 
\begin{equation}\label{BC gauge}
\mathcal{A}_1|_{\partial\mathcal{SM}^{(3|4)}}= \widetilde{B} \wedge \mathbb{Y}^{(0|2)}, \qquad \mathcal{A}_2|_{\partial \mathcal{SM}^{(3|4)}}=  B 
\end{equation}
which ensure that the boundary term arising from the gauge variation of $S_{\rm sBF}$ in \eqref{susycompacglobal} gets exactly canceled at the physical boundary by the anomalous variation \eqref{hoE} of the physical theory. \\
The PCO $\mathbb{Y}^{(0|2)}$ necessarily introduced in \eqref{BC gauge} for matching the form picture, localizes the expression on two out of four fermionic directions. Therefore, it breaks ${\cal N}=2$ supersymmetry in three dimensions (four real supercharges) to ${\cal N}=1$ in two dimensions (two real supercharges). This is how in supergeometry we formalize the fact that the presence of a boundary halves supersymmetry. 

However, this is not the end of the story, as we have also to investigate how the bulk action \eqref{susycompacglobal} behaves under a supersymmetry transformation in the bulk. 
As we discuss in detail in appendix \ref{appD},
the supersymmetry variation produces a boundary term, which has to vanish in order to restore supersymmetry. Remarkably, the boundary conditions in \eqref{BC gauge}, required for matching the t'Hooft anomaly at the boundary, are also sufficient for canceling the boundary term coming from the supersymmetry variation of $S_{\text{sBF}}$. Therefore, the full supersymmetry invariance in the bulk is preserved, though at the price of imposing half-preserving boundary conditions.

At the topological boundary, we fix an absolute theory by either fixing a polarization for the $\mathbb R \times \mathbb R$ symmetry group or by coupling to topological edge modes. We will take here the former approach.  To retrieve the super-compact boson, we specify a Lagrangian submanifold by picking a maximal torus $(\mathbb R /2\pi \mathbb{Z})^{(0|0)} \times (\mathbb R /2\pi\mathbb{Z})^{(0|2)}$ of mutually local operators at the boundary (this maximal torus trivializes the braiding phase \eqref{nonchibraiding}). 
Different choices of boundary conditions are associated to distinct theories at the physical boundaries. Fixing a maximal torus for which $\mathcal{A}_2|_{\mathcal{SM}_{\rm top}^{(2|2)}}=0$, realizes a theory with a $\mathbb{R}$ ${(0|0)}$-global symmetry, and vice-versa, $\mathcal{A}_1|_{\mathcal{SM}_{\rm top}^{(2|2)}}=0$ one with  $\mathbb{R}$ ${(0|2)}$-global symmetry.

We close this section by discussing the reduction to 
components of  $S_{\text{sBF}}$ in \eqref{susycompacglobal}, in particular how to link it to the conventional ${\mathcal N}=1$ supersymmetric BF action \cite{Gates:1983nr}. 

To this end, it is convenient to re-express \eqref{susycompacglobal} as follows 
\begin{equation}
    \label{tqB}
     \int_{\mathcal{M}^{(3|4)}} \mathcal{L}^{(3|0)} \wedge \mathbb{Y}^{(0|4)}
\end{equation}
where the PCO is given by 
\begin{equation}
    \label{tqC}
    \mathbb{Y}^{(0|4)} = (V\wedge V)^a (\theta_1^2 \iota_2\gamma_a \iota_2 + \theta_2^2 
    \iota_1\gamma_a \iota_1) \delta^2(\psi_1) \delta^2(\psi_2) 
\end{equation}
This expression is obtained by the supergeometry setting in D=4 $\mathcal N=1$ framework and it shows a phenomena discussed in \cite{Castellani:2015ata} 
leading to the chiral and antichiral splitting of supersymmetric actions 
(a complete discussion is found also 
in \cite{Castellani:2023tip}). A remark is needed: the expression of the PCO in \eqref{tqC} is not manifestly supersymmetric, but it susy variation is $d$-exact, and since the Lagrangian $\mathcal{L}^{(3|0)}$ is closed 
this fact is harmless. 

The CS $\mathcal{N }=2$ theory in geometric formalism was obtained in 
\cite{Grassi:2016apf,Cremonini:2019aao,Fre:2017aqt} and using the PCO \eqref{tqC} the CS action is divided in two pieces the $\mathcal{N}=1$ CS and a mass term for a Wess-Zumino multiplet.

\section{Chiral Supermultiplet}

We now consider another simple example, that is the model describing the free dynamics of a chiral supermultiplet in two dimensional $N=(1,1)$ superspace. In the present context, this is described by a $(0|0)$-superform $\Phi$ satisfying $\overline{D} \Phi=0$ (and then  $\overline\partial \Phi=0$). We will rely on notations and conventions used in the previous sections, with the addition of this extra condition. 

The chirality constraint implies that the degrees of freedom reduce to a single holomorphic boson and a single holomorphic fermion. In fact, given the general expansion \eqref{eq:exp} for the $\Phi$ superfield, the chirality constraint reduces it to 
\begin{equation}
\Phi = \phi + \theta  \lambda \, ,  \qquad \quad W \equiv  D\Phi = \lambda - \theta \partial \phi
\end{equation}
with $\overline\partial \phi = \overline\partial \lambda = 0$. The action of the model has been written in \cite{Sen:2015nph,Sen:2019qit,Hull:2023dgp}, and in \cite{Cremonini:2020skt} within the framework of supergeometry. 

\medskip
The results found above for the chiral sector of the compact boson can be easily adapted to the present model. However, as we are going to discuss, this model exhibits two additional  features. First of all, compared to the previous example, it possesses only one conserved U(1) current, due to its self-duality. As we are going to show, this current is not anomalous and does not exhibit any mixed `t Hooft anomaly with the supersymmetry current.  
Moreover, the model suffers from a gravitational anomaly, due to the presence of a chiral multiplet. The gravitational anomaly has been computed long ago in \cite{Alvarez-Gaume:1983ihn}.
Here, we look for the structure of the anomaly inflow that can be introduced to compensate it, formulated in the language of supergeometry. This extra term modifies the form of the  symTFT which is no longer represented simply by the BF model associated with the mixed `t Hooft anomaly. 

\subsection{\texorpdfstring{Self-duality of the $U(1)$ currents}{Self-duality of the U(1) currents}}

In this section we prove that the chirality constraint $\overline{D} \Phi =0$ implies self-duality of the $U(1)$ current $J = d\Phi$. To this end, we need first to introduce the definition of self-duality in supermanifolds. 

In conventional bosonic manifolds, the (anti-)self-duality of a two-dimensional boson is expressed by the condition 
\begin{equation}
    \label{sdA}
     \star d \phi = \pm\, d\phi  .
\end{equation}
Bianchi identity then implies the equation of motion $d\star d\phi=0$. Conditions \eqref{sdA} are satisfied for $\overline\partial \phi = 0$ (or $\partial \phi = 0$ in the anti-self-dual case). 

In a supermanifold, the situation is more involved since we have to take into account the presence of  fermions in the multiplet. 
In this case, the (anti-)self-duality condition \eqref{sdA} is promoted to an analogue condition for the superform $\Phi$ \cite{Cremonini:2020skt} 
\begin{equation}
    \label{sd2A}
     \star d\Phi = \pm d\Phi \wedge \mathbb{Y}^{(0|2)} 
\end{equation}
where the PCO $\mathbb{Y}^{(0|2)}$ has been inserted to match the order and the picture of the two integral/super-forms on the two sides. Applying one extra $d$ operator, this identity leads to the equations of motion $d(\star d\Phi)=0$. 

Alternatively, the self-duality condition can be   written as
\begin{equation}
    \label{sd2BA}
    \mathbb{Z}\overline{\mathbb{Z}}[\star d\Phi] = \pm d\Phi 
\end{equation}
where $\mathbb{Z}, \overline{\mathbb{Z}}$ are the two lowering PCOs introduced in appendix \ref{sec:PCO}. 

Upon choosing the following PCO\footnote{ This PCO is closed, but not exact, and differs from the more standard $\mathbb{Y}^{(0|2)} = \theta\overline\theta \delta(\psi) \delta(\overline\psi)$ by exact terms.}
\begin{equation}
    \label{sd2B}
    \mathbb{Y}^{(0|2)} = \delta^2(\psi) - V \overline{V} \delta'(\psi) \delta'(\overline\psi)
\end{equation}
and expanding $d\Phi$ and $\star d\Phi$ as in \eqref{hoB}, constraint \eqref{sd2A} enforces and is enforced by the usual (anti)-self-duality condition $\overline{D} \Phi = 0$ (or $D\Phi = 0$ in the anti-self-dual case), which is nothing but the chiral constraint. 

Therefore, we can conclude that for a chiral multiplet the two conserved  currents $J = d\Phi$ and $\tilde{J} = \star d\Phi$ satisfy the self-duality identity $J = \tilde{J}$, thus they describe the same $U(1)^{(0|0)}$ symmetry. 
Since supersymmetry and $\kappa$-symmetry are diffeomorphisms of the supermanifold, the self-duality in \eqref{sd2A} easily implies the same self-duality condition for ${\mathcal J} = {\mathcal L}_\epsilon J$ and ${\mathcal J}_{\kappa} = {\mathcal L}_\kappa J$ introduced in \eqref{eq:susytransformedcurrents} and \eqref{Kcurrents}, respectively. 

The chirality - or equivalently the self-duality - constraint implies that the  expansion of the  $(1|0)$-supercurrent $J$ is shortened to 
\begin{equation}
 J = V \partial \Phi + \psi D\Phi  
\end{equation}
with holomorphic components, $\overline{D} (\partial \Phi) = \overline{D} (D\Phi)=0$ (then $\overline\partial (\partial \Phi) = \overline\partial (D\Phi)=0$).
Generalizing, any self-dual $(1|0)$-superform has a chiral expansion of this form.

\subsection{Mixed and gravitational anomalies}
We now study the possible emergence of 't Hooft anomalies associated to the global symmetry, and mixed gauge-supersymmetry anomalies, in analogy with the previous example.   

We first consider coupling the theory to a self-dual, chiral background \begin{equation}
    B = b_V V + b_\psi \psi
\end{equation}
where $B$ is a $(1|0)$-superform with $\overline{D} b_V = \overline{D} b_\psi = 0$ (and then $\overline\partial b_V = \overline\partial b_\psi =0$).\footnote{We note that the holomorphy conditions together with the conventional constraints on the field-strength lead to $dB=0$.} We use $B$ to covariantize $J$ with respect to the chiral shift, $\Phi \to \Phi + \omega, B \to B + d\omega$, with $\overline{D} \omega = 0$.

The supersymmetry current is given by the chiral component of one of the currents (\ref{sussaA}-\ref{sussaB}), according to the choice of the form picture. As in the previous example, we can use the $B$ components to covariantize ${\mathcal S}^{(1|1)}$ respect to the $U(1)$ shift symmetry, obtaining ${\mathcal S}_{\rm cov}$ in eq. \eqref{covcurrents}. 
We then consider the linear couplings with a chiral $U(1)$ and a gravitino backgrounds  
\begin{equation}
S_{\rm linear}= \int_{\mathcal{SM}^{(2|2)}} \Big( (d\Phi - B) \wedge \widetilde{B} + \chi {\mathcal S}_{\rm cov}\Big)
\end{equation}
and study its variation 
under gauge transformations $B \to B + d\omega$,  $\widetilde{B} \to \widetilde{B} + d\widetilde\omega$ and $\chi \to \chi + d\Lambda$. 
By construction, the action is manifestly invariant under $B$ transformations. It is also invariant under $\widetilde{B}$ transformations because of the chirality of $(d\Phi -B)$, whereas under $\chi$ variations, we obtain 
\begin{equation}
\int_{\mathcal{SM}^{(2|2)}} \Lambda \,d{\mathcal S}_{\rm cov} = \int_{\mathcal{SM}^{(2|2)}} \Lambda \, \big[db_\psi(\partial\Phi-b_V)+d b_V(W-b_\psi)\big]V\delta(\psi)
\end{equation}
However, also this expression vanishes, thanks to the chirality of $B$.
We then conclude that in the chiral model there is no mixed gauge-supersymmetry anomaly. 

\medskip

This does not exhaust the spectrum of anomalies of the chiral boson. In fact, it is well known that this model possesses a gravitational anomaly. 
In order to take it into account, we linearly couple the supersymmetry current $\mathcal{S}^{(1|1)}$ to a gravitational background field $H^{(1|1)}$. 

A generic $(1|1)$-pseudoform has the following structure 
\begin{equation}
\begin{split}
    \label{muF}
&\hspace{-1.0cm}  H^{(1|1)} = 
               \sum_{p\geq 0} A_p \psi^{p+1} \delta^{(p)}(\overline \psi) 
              + \sum_{p\geq 0} \overline A_p \overline\psi^{p+1} \delta^{(p)}(\psi)   
              +
               \sum_{p\geq 0} B_p V \psi^p \delta^{(p)}(\overline \psi) 
              + \sum_{p\geq 0} \overline B_p \overline V \psi^p \delta^{(p)}(\overline \psi)
               \\
              & \hspace{-1.0cm} +
               \sum_{p\geq 0}  C_p  V \overline\psi^p \delta^{(p)}(\psi) 
              + \sum_{p\geq 0} \overline C_p  \overline V \overline\psi^p \delta^{(p)}(\psi)
              +
              \sum_{p\geq 0} D_p V\overline V \psi^p \delta^{(p+1)}(\overline \psi) 
              + \sum_{p\geq 0} \overline D_p V \overline V \overline\psi^p \delta^{(p+1)}(\psi) 
\end{split}
\end{equation}
where the coefficients of the expansion are superfields. Here, the upper indices of the delta's denote the order of their derivatives w.r.t. $\psi$ or $\overline \psi$. Powers of $\psi$ or $\overline{\psi}$ are intentionally chosen to compensate the negative form degrees of $\delta^{(p)}(\overline \psi), \delta^{(p)}(\psi)$.  

The $H^{(1|1)}$ pseudoform is defined up to the following infinite chain of gauge transformations
\begin{equation}
    \label{muFA}
    \delta  H^{(1|1)} = d \Sigma^{(0|1)}\,, ~
    \delta \Sigma^{(-p|1)} = \Sigma^{(-p-1|1)}\,, ~
    \forall p \geq 0  
\end{equation}
for generic pseudoform parameters $\Sigma$. 

Evaluating the linear coupling between the supersymmetry current and the background, we obtain 
\begin{equation}
    \label{muG}
    \int_{{\mathcal SM}^{(2|2)}} \!\!  \mathcal{S}^{(1|1)} \wedge H^{(1|1)} 
    = 
     \int_{{\mathcal SM}^{(2|2)}} \!\! (W 
    \partial \Phi) V \delta(\psi) \wedge \overline B_0 \overline V \delta(\overline\psi) 
    = 
     \int d^2z  \, d^2\theta \,  (W \partial\Phi)  \wedge \overline B_0 
\end{equation}
The reason why only the $\overline{B}_0$ component of the expansion \eqref{muF} survives can be traced back to the anticommuting nature of $V, \overline V$ and the obvious identities 
$\overline\psi \delta(\overline\psi) = 0$ and $\delta(\psi) \delta(\psi) =0$. \\
The action in \eqref{muG} is manifestly invariant under gauge transformations \eqref{muFA} thanks to the closure of $\mathcal{S}^{(1|1)}$. 

In order to perform the Berezin integral in \eqref{muG} and obtain the gravitational coupling in components, it is 
convenient to work in the Wess-Zumino gauge, $\overline \theta \overline B_0 =0$. This is equivalent to choosing
\begin{equation}
    \label{muH}
    \overline B_0 = \overline \theta( h +  \theta\chi   ) 
\end{equation}
where $h$ and $\chi$ are bosonic and fermionic ordinary fields, respectively. Inserting this expansion together with the one for $W\partial \Phi$ (see eq. \eqref{muD}) and evaluating the Berezin integral, we eventually obtain
\begin{equation}
    \label{muI} 
    \int_{\mathcal{SM}^{(2|2)}} \!\!   \mathcal{S}^{(1|1)} \wedge H^{(1|1)} 
=\int d^2 z
 \left[ \lambda \partial \phi \, \chi + (\lambda \partial \lambda + (\partial \phi)^2 ) \, h \right]  
\end{equation}
We recognize $h$ playing the role of the vielbein coupled to the stress-energy tensor and $\chi$ the gravitino coupled to the supersymmetry current. 
Given tranformations \eqref{muE} for the current components, the coupling is supersymmetric if 
\begin{equation}
    \label{muIA}
    \delta_\epsilon h = \epsilon \chi\,, ~~~~~~
    \delta_\epsilon \chi = - \partial(\epsilon h)
\end{equation}
These are the well known supergravity transformation rules with local supersymmetry parameter $\epsilon$. 

The linear coupling \eqref{muI} leads to an anomalous Ward identity \cite{Howe:1985uy,Alvarez-Gaume:1983ihn}  which represents the easiest example of gravitational anomaly for the chiral supermultiplet. 
To account for this anomaly the SuSymTFT needs to be complemented by a Chern-Simons term, which in the present context reads 
\begin{equation}
    \label{muL}
    S_{CS} = \alpha \int_{\mathcal{SM}^{(3|4)}} H^{(1|2)} \wedge d H^{(1|2)}
\end{equation}

The $\alpha$ coefficient has been determined in \cite{Howe:1985uy} by evaluating the supersymmetry variation of the one-loop effective action. It is subject to suitable quantization conditions to ensure invariance under background diffeomorphisms (see \cite{Chang:2020aww} for a general discussion). 

The relation between the integral forms $H^{(1|2)}$ and the boundary $H^{(1|1)}$ is via a suitable gauge fixing of the extra $\theta$-coordinates. As previously described, this can be done by using the PCO along the additional fermionic variables, without affecting the physics. 

Note that the expansion of $H^{(1|2)}$ differs from 
\eqref{muF} since the supermanifold has acquired a new bosonic direction. This construction has been studied in \cite{Grassi:2016apf, Cremonini:2019aao, Cremonini:2019xco} and for 3d supergravity in \cite{Castellani:2020kmz}.

\section{Outlook}

In this paper we have proposed an intrinsically supersymmetric formulation of the SymTFT -- the SuSymTFT. By formulating  the SymTFT directly into superspace, one gains  the advantage of organizing the symmetries and anomalies of a supersymmetric theory directly into their natural multiplets conveniently packaged in superfields. In this work  we have illustrated the construction explicitly in the case of two-dimensional (chiral and non chiral) $\mathcal{N} =(1,1)$  theories. 

We have advanced a formal construction of the SymTFT in a supersymmetric environment, but more work is required to embed our construction within the most general systematic study of spacetime and superconformal symmetries \cite{Apruzzi:2025SpacetimeSymTFT}. Furthermore, the SuSymTFT has to be better investigated close to the symmetry boundary by constructing the quantum algebra of topological and charged operators. 

\vskip 5pt
\noindent
This work leaves many open interesting direction for future investigations:
\begin{itemize}
\item Our  construction could be straightforwardly generalized to the chiral ${\cal N}=(2,0)$ superspace.
\item We have considered the case of minimal supersymmetry in two dimensions. To do that, we have placed the physical theory on a 1/2-BPS boundary of the bulk supermanifold $\mathcal{M}^{(3|4)}$ supporting an $\mathcal{N}=2$ topological theory.  It would be interesting to explore more general cases where already the bulk exhibits extended supersymmetry.
    \item Here we limited ourselves to constructing explicitly the SuSymTFT for the global symmetries.  Following the recent construction of a the SymTFT for spacetime symmetries \cite{Apruzzi:2025SpacetimeSymTFT}, it would be very natural to also include the global supersymmetry algebra in such a construction. This would naturally entail coupling the bulk to (topological) supergravity. In this direction, we have done a  preliminary discussion in the case of the chiral boson, but a systematic investigation is still missing.
    \item While we have presented our proposal in full generality, as a first application we have considered only the case of two-dimensional theories. It would  then be natural to investigate the application of our construction to the case of supersymmetric theories in higher dimensions, where the structure of superspace, as well as the choices of boundary conditions and the selection of topological bulk theories would be much richer. 
    \item As an aside problem, it would be interesting to investigate the alternative construction of half-BPS boundary supersymmetric theories without boundary conditions, exploiting the F+A construction of \cite{Belyaev:2008xk} rephrased in the supergeometry language. It would be interesting to understand whether such a construction may play a role in the Symmetry TFT approach.
\end{itemize}

\section*{Acknowledgments}

PAG would like to thank Carlo Alberto Cremonini, Ruggero Noris and Lucrezia Ravera for useful discussions on super-BF theories. FA thanks Yi Zhang for useful discussions, and the Physics Department of the University of Milano-Bicocca for the hospitality during which part of the research for this work was performed. PAG would like to thank the CERN TH-Department where this works has been done. SP would like to thank the Isaac Newton Institute for Mathematical Sciences, Cambridge, for support and hospitality during the programme "Quantum Field Theory with Boundaries, Impurities, and Defects" where work on this paper was undertaken. This work was supported by EPSRC grant no EP/R014604/1. Research at Perimeter Institute is supported in part by the
Government of Canada through the Department of Innovation, Science and Economic Development Canada and
by the Province of Ontario through the Ministry of Colleges and Universities.
PAG is partially supported by HORIZON-MSCA-2021-SE-01-101086123 CaLIGOLA and
by INFN grant Gauge Theory, Supergravity, Strings (GSS 2.0). AD and SP are partially
supported by the INFN grant Gauge and String Theory (GAST). 

\appendix

\section{Picture Changing Operators: A  primer}\label{sec:PCO}

Here we depict a few ingredients regarding PCOs, which we use in the main text. For the general construction of PCOs in supergeometry we refer to \cite{Belopolsky:1997jz}, whereas for an exhaustive review on PCOs in supermanifolds see \cite{Grassi:2025tfp}. 

As in string theory, there two types of PCO's: picture raising operators and picture lowering operators. In the present context the picture number is given by the number of Dirac delta's function $\delta(\psi)$ in a given expression, where $\psi$ is the odd vielbein ($\psi = d \theta$ in flat superspace). Lowering means removing delta's, raising means increasing the number of delta's. 

We recall that a raising PCO $\mathbb{Y}^{(0|q)}$ (increasing the picture by $q$ units, without changing the form degree) has the following properties 
\begin{equation}
\begin{split}
    \label{PCOA}
    &d \mathbb{Y}^{(0|q)} =0\,, ~~~~~~~
    \mathbb{Y}^{(0|q)} \neq d K^{(-1|q)}\,, ~~~~~
     \\
    &
    \omega^{(r|s)} \rightarrow  \omega^{(r|s+q)} =  \omega^{(r|s)}\wedge \mathbb{Y}^{(0|q)}
\end{split}
\end{equation}
where $ \omega^{(r|s)}$ is generic $(r|s)$ pseudoform. 

The PCO is an isomorphism on the cohomology of $d$, namely $H^{(r|s)}$. Since the PCO is a representative of the cohomology class
$H^{(0|q)}$, one can choose different but equivalent expressions. For instance, in a ${\mathcal{SM}}^{(2|2)}$ supermanifold with $(z, \overline{z}, \theta, \overline\theta)$ coordinates (see appendix \ref{appA} for details), taking $q=2$ we can consider the following different PCO realizations  
\begin{equation}
\begin{split}
    \label{PCOB}
    \mathbb{Y}^{(0|2)} &= \theta \overline \theta \delta(\psi) \delta(\overline \psi)\,, ~~~~~  \\
    \mathbb{Y}^{(0|2)} &= V\delta'(\psi) \overline V \delta'(\overline \psi)
    \,, ~~~~~  \\
    \mathbb{Y}^{(0|2)} &= \delta(\psi) \delta(\overline\psi) + V\delta'(\psi) \overline V \delta'(\overline \psi)
    \end{split}
\end{equation}
where $V, \overline{V}$ are the bosonic vielbeins and $\delta'(\psi) \equiv \frac{\partial}{\partial \psi} \delta(\psi)$ stands for $\iota_D \delta(\psi)$, $\iota_D$ being the the contraction operator (inner derivation) along the supersymmetry covariant derivative $D$. The first expression is not manifestly supersymmetric, whereas the second and the third ones are manifestly supersymmetry invariant. The first and the second PCOs have a non-trivial kernel, the last one has no kernel. 

Lowering PCOs, denoted as 
$\mathbb{Z}^{(0|-q)}$, instead remove $q$ delta's. They have the following properties 
\begin{equation}
\begin{split}
    \label{PCOC}
    &d \mathbb{Z}^{(0|-q)} =0\,, ~~~~~~~
    \mathbb{Z}^{(0|-q)} \neq d K^{(-1|-q)}\,, ~~~~~
     \\
    &
    \omega^{(r|s)} \rightarrow  \omega^{(r|s-q)} =  \mathbb{Z}^{(0|-q)}[\omega^{(r|s)}]
   \end{split}
    \end{equation}
with $s \geq q$. 
Again, these PCOs are isomorphisms in the cohomology. An explicit realization for $q=-1$ is for example \cite{Belopolsky:1997jz}
\begin{equation}
    \label{sdD}
     \hspace{-.4cm}\mathbb{Z}^{(0|-1)} = \Big[d, \Theta(\iota_{D})\Big] = \mathcal{L}_{D} \delta(\iota_{D}) + \iota_{\overline\partial} \delta'(\iota_{D})
\end{equation}
where $\Theta$ is the Heaviside theta function, ${\cal L}_D$ and $\iota_{D}$ are the Lie derivative and  the contraction operator along the vector field $D$, respectively, while 
$\iota_{\partial}$ is the contraction along the vector field $\partial$. 
The expression $\delta(\iota_{D})$ is the Dirac delta of the contraction operator. It acts on $\delta(\psi)$ as 
$\delta(\iota_{D})\delta(\psi)=1$ (up to normalization factors). 

Given  $\mathbb{Z}^{(0|-1)}$ in \eqref{sdD} we have the freedom to add exact terms without modifying the cohomology class. It follows that expression \eqref{sdD} is supersymmetric and ``almost'' universal, 
in the sense that the super-derivative $D$  can always be replaced by any other odd vector field $X$, having the replacement the only effect of generating extra exact terms. 

Finally, we note that being $\mathbb{Z}^{(0|-1)}$  expressed in terms of the Heaviside theta function - which is not a compact-support distribution - it is an element of the cohomology, but as a representative is defined up to exact terms.

In general, the following identities hold
\begin{equation}
    \label{PCOD}
    \mathbb{Z}^{(0|-1)} \mathbb{Y}^{(0|1)} \mathbb{Z}^{(0|-1)}= \mathbb{Z}^{(0|-1)}\,, \qquad \quad 
    \mathbb{Y}^{(0|1)} \mathbb{Z}^{(0|-1)} \mathbb{Y}^{(0|1)}= \mathbb{Y}^{(0|-1)} ~~~~~~
\end{equation}
and for particular choices of the odd vector field in the definition of the PCO, one can find the inverse $ \mathbb{Z}^{(0|-1)} \mathbb{Y}^{(0|1)} =1$ in the full space of superforms or integral forms.

\section{\texorpdfstring{The $\mathcal{SM}^{(2|2)}$ supermanifold}{The SM(2|2) supermanifold}}
\label{appA}

We parametrize the  $\mathcal{SM}^{(2|2)}$ minkowskian superspace  by two bosonic light-cone real coordinates $(x^{++} , x^{--} ) \equiv (z, \overline{z})$ and one Majorana spinor $\theta^\alpha = (\theta^+, \theta^-) \equiv (\theta, \overline\theta)$ with real entries. 

The corresponding differential $1$-forms are 
$(dz, d\overline z, d\theta, d\overline\theta)$ and the flat super-vielbeins read
\begin{equation}
\label{oneA}
V = dz + \theta d\theta \, ,
\quad \overline V= d\overline z + \overline \theta d\overline\theta \, ,  \qquad \psi = d\theta \, , \quad
\overline\psi = d \overline \theta
\end{equation}
They are supersymmetric invariant under $\delta \theta = \epsilon, \delta \overline{\theta} = \overline{\epsilon}$ and $\delta z = \epsilon \theta, \delta \overline{z} = \overline{\epsilon} \overline{\theta}$. In fact, for rigid supersymmetry ($\epsilon, \overline{\epsilon}$ constant) 
\begin{equation}
\begin{split}
	\delta V &= d \left( \delta z \right) + \delta \theta d \theta + \theta d \left( \delta \theta \right) =  - \epsilon d \theta + \epsilon d \theta + \theta d \epsilon = 0   \\
	\delta \psi &= d \left( \delta \theta \right) = d \epsilon = 0
    \end{split}
\end{equation}
The super-zweibeins in \eqref{oneA} satisfy the Maurer-Cartan equations 
\begin{equation}
\label{oneAB}
d V = \psi\wedge \psi\,, \quad 
d \overline V = \overline\psi\wedge \overline\psi\,, \qquad
d\psi =
d\overline\psi =0
\end{equation}

If we set $V^a \equiv (V, \overline{V})$ and $\psi^\alpha \equiv (\psi, \overline\psi)$, the action of the super-Hodge dual reads \cite{Castellani:2015dis}
\begin{equation}
    \star V^a  =  \epsilon^{ab} \, V_b \, \delta^{(2)} (\psi) \, , \qquad 
    \star \psi^\alpha = V \overline{V} \epsilon^{\alpha \beta} \iota_\beta \, \delta^{(2)} (\psi)
\end{equation}
where we have defined the antisymmetric tensors $\epsilon^{z\bar{z}} = \epsilon_{\bar{z}z} =2, \epsilon^{+-} = \epsilon_{-+} = 1$, and $\delta^{(2)} (\psi) \equiv \delta (\psi)\delta (\overline\psi)$.

Covariant derivatives in superspace are defined as
\begin{equation}\label{algebra2d}
\partial_z \equiv \partial \, ,  \quad \partial_{\overline z} \equiv \overline{\partial} \, , \qquad
	D = \partial_\theta -  \theta \partial \, , \quad \overline D = \partial_{\overline\theta} - \overline\theta \overline\partial
\end{equation}	
They form a 
	representation of the Lie superalgebra 
	\begin{equation}
\label{oneABA}
D^2 = - \partial \ , \ \overline D^2 = - \overline\partial \ , \qquad
\left\{ D , \overline{D} \right\} =0 
\end{equation}

Correspondingly, supersymmetry charges are given by
\begin{equation}\label{algebra2d_charges}
	Q = \partial_\theta +  \theta \partial \ , \ \overline Q = \partial_{\overline\theta} + \overline\theta \overline\partial \ , 
\end{equation}	
and satisfy the following superalgebra 
\begin{eqnarray}
\label{oneABAQ}
&&\ Q^2 = \partial \ , \ \overline Q^2 =  \overline\partial \ , \quad
\left\{Q , \overline{Q} \right\} =0 \,, 
\nonumber \\ 
&&\{D, Q\} = \{D, \overline Q\} = \{\overline D, Q\} = 
\{\overline D, \overline Q\} =0 
\nonumber 
\end{eqnarray}

In terms of superspace covariant derivatives, the differential acting on superfields is defined as
\begin{equation}\label{eq:d}
   d = V \partial + \overline{V} \overline{\partial} + \psi D + \overline{\psi} \overline{D} \, .
\end{equation}

According to the general construction, in ${\cal{SM}}^{(2|2)}$ we can define $\omega^{(p|q)}$ pseudoforms at different pictures, with $q=0,1,2$. A zero-picture form  $\omega^{(p|0)}$ has an expansion in terms of $dz, d\overline z, d\theta, d\overline\theta$ differentials, while the expansion of a  maximal-picture form 
 $\omega^{(p|2)}$ is expressed in terms of 
 $dz, d\overline z$ and the product of Dirac delta's   $\delta(\psi), \delta(\overline\psi)$ and derivatives thereof. In the main text, the expansions are made explicit whenever this is instrumental for the comprehension.

\section{Mixed anomaly for the compact boson: The details}\label{app:mix-anomaly-details}

In this appendix we detail the discussion of the gauge invariance of the action in \eqref{hoD} for the super-compact boson and determine explicitly the background gauge transformations that leave the action invariant, up to the  't Hooft anomaly \eqref{hoE}.
 
In \eqref{hoD}, let us focus on the coupling of the supersymmetry current to the gravitino background 
\begin{equation}\label{linearcoupling}
   \int_{\mathcal{SM}^{(2|2)}} \chi \wedge( {\mathcal S}_{\rm cov} +  \overline {\mathcal S}_{\rm cov} )
    \end{equation}
   where ${\mathcal S}, \overline{\mathcal S}$ are given in \eqref{covcurrents}.  
    If we now perform a gauge transformation $\chi\rightarrow \chi+d\Lambda$ with arbitrary $(0|1)$-form gauge parameter $\Lambda$, the action varies by the following expression
\begin{equation}\label{GAnomaly}
       -\Lambda\left\{\big[db_\psi(\partial\Phi-b_V)+d b_V(W-b_\psi)\big]V\delta(\psi)+\left[d\overline b_\psi(\overline\partial\Phi-\overline{b}_V)+d\overline{b}_V(\overline W-\overline b_\psi)\right]\overline V\delta(\bar \psi)\right\} 
    \end{equation}
integrated on ${\mathcal SM}^{(2|2)}$.
The most general expansion of $\Lambda$ in components is given by an infinite sum,
\begin{eqnarray}
&&\hspace{-1.0cm}  \Lambda = 
               \sum_{p\geq 0} \bar a_p \psi^{p} \delta^{(p)}(\overline \psi) 
              + \sum_{p\geq 0}  a_p \overline\psi^{p} \delta^{(p)}(\psi)   
              +\sum_{p\geq 0}  b_p  V \overline\psi^p \delta^{(p+1)}(\psi)
              + \sum_{p\geq 0} \overline {b_p V} \psi^p \delta^{(p+1)}(\overline \psi)
              \nonumber \\
              && \hspace{-1.0cm} +
                \sum_{p\geq 0} c_p V \psi^p \delta^{(p+1)}(\overline \psi)
              + \sum_{p\geq 0} \overline {c_p V }\overline\psi^p \delta^{(p)}(\psi)
              +
              \sum_{p\geq 0} d_p V\overline V \psi^p \delta^{(p+2)}(\overline \psi) 
              + \sum_{p\geq 0} \overline d_p V \overline V \overline\psi^p \delta^{(p+2)}(\psi)  \nonumber \\
\end{eqnarray}
where the coefficients are ordinary superfields and $\delta^{(p)}(\psi)$ means the $p$-derivative of the delta function. 
However, when $\Lambda$ appears as in \eqref{GAnomaly} multiplied by terms proportional to $V\delta(\psi)$ and $\overline V\delta(\bar\psi)$, only a finite number of terms in the expansion survive. Precisely, 
expression \eqref{GAnomaly} reduces to
\begin{equation}\label{GAfact}
    -\overline\lambda\big[db_\psi(\partial\Phi-b_V)+d b_V(W-b_\psi)\big]V\delta(\psi)-\lambda\left[d\bar b_\psi(\bar\partial\Phi-\bar{b}_V)+d\bar{b}_V(\overline W-\bar b_\psi)\right]\overline V\delta(\bar \psi)
\end{equation}
where we have defined
\begin{equation}  \lambda=a_0\delta(\psi)+b_0 V\delta'(\psi) ~~~,~~~ \overline\lambda=\overline a_0\delta(\bar\psi)+\overline {b_0  V}\delta'(\bar\psi) 
\end{equation}

Now, it is easy to see that expression \eqref{GAfact} can be recast in a very compact form, as
\begin{equation}
\Sigma(\Lambda,B) \wedge(d\Phi-B)  
\end{equation}
where $\Sigma(\Lambda,B)$ is the $(1|2)$-superform 
\begin{equation}  \label{sigma}  \Sigma(\Lambda,B) =\bar\sigma^{(1|1)}\delta(\psi)+\bar\tau^{(1|1)}V\delta'(\psi)+\sigma^{(1|1)}\delta(\bar\psi)+\tau^{(1|1)}\overline V\delta'(\bar \psi)
\end{equation}
whose form components read
\begin{equation}
\begin{split}   
\label{sigmacomponents}
\bar\sigma^{(1|1)}&=\bar\lambda \, (\overline V \overline \partial b_\psi + \overline \psi \overline D b_\psi) ~~~,~~~ \sigma^{(1|1)}=\lambda \, ( V  \partial \overline b_\psi +  \psi  D \overline b_\psi)\\
\bar\tau^{(1|1)}&=\bar\lambda \, (\overline V \overline\partial b_V + \overline \psi \overline D b_V) ~~~,~~~ \tau^{(1|1)}=\lambda \, ( V  \partial\overline b_V +  \psi D \overline b_V)  
\end{split}
\end{equation}
In conclusion, we have found that under a gravitino gauge transformation featured by parameter $\Lambda$ the gravity part of the action varies as
\begin{equation}\
   \delta \int_{\mathcal{SM}^{(2|2)}} \chi \wedge( {\mathcal S}_{\rm cov} +  \overline {\mathcal S}_{\rm cov} ) = \int_{\mathcal{SM}^{(2|2)}} \Sigma(B, \Lambda)\wedge(d\Phi-B)
    \end{equation}
It is then straightforward to see that this variation can be canceled by modifying the gauge transformation of $\widetilde{B}$ as 
\begin{equation}
    \widetilde{B}\rightarrow \widetilde{B}+d\widetilde\omega -\Sigma(\Lambda,B)
\end{equation}
so that, referring to eq. \eqref{hoD} we obtain
\begin{eqnarray}
    \delta S[B,\widetilde{B}, \chi ] &=& \frac{1}{4\pi} \int_{{\mathcal SM}^{(2|2)}} \Big[(d \widetilde\omega - \Sigma(\Lambda, B) ) \wedge (d\Phi-B) + \Sigma(\Lambda, B)  \wedge (d\Phi-B) \Big] \nonumber \\
    &=& \frac{1}{4\pi} \int_{{\mathcal SM}^{(2|2)}} \widetilde\omega \wedge dB
\end{eqnarray}
This is the expected mixed anomaly that we discussed in the main text. 

\section{Supersymmetry with boundary}\label{appD}

In this appendix we discuss the supersymmetry invariance of the SuSymTFT action for the super-compact boson introduced in section \ref{sect:symTFTcompactboson}.

In supergeometry, an infinitesimal supersymmetry transformation is captured by the Lie derivative with respect to the vector field $\epsilon=\epsilon^\alpha Q_\alpha$, where $Q_\alpha$ are the real supercharges of the theory. 

The integral of a top form $\omega$ on a manifold without boundary is automatically supersymmetric, as 
\begin{equation}
\label{AppD1}
    \int_\mathcal{SM} \mathcal{L}_\epsilon\omega \equiv \int_\mathcal{SM} \iota_\epsilon (d\omega)+d(\iota_\epsilon\omega)=0
\end{equation}
In fact, the first term vanishes because $\omega$ is a top form, while the second term vanishes by applying Stokes theorem in the absence of a boundary. 

Instead, if the supermanifold has a boundary the last term in \eqref{AppD1} survives as a boundary term
\begin{equation}\label{bdryD}
\int_{\partial\mathcal{SM}}\iota_\epsilon\omega
\end{equation}

This term has to be removed in some way in order to restore supersymmetry.
One possibility would be  adding a boundary action whose variation cancels exactly (\ref{bdryD}). This would preserve supersymmetry without imposing any boundary condition. However, in the present context this approach is not necessary since in any case we have to impose boundary conditions to match the boundary t'Hooft anomaly. 
In fact, in this appendix we show that boundary conditions \eqref{BC gauge} on the bulk fields are already sufficient to make the supersymmetry variation of $S_{\text{sBF}}$ in \eqref{susycompacglobal} vanish, upon a suitable choice of the PCO.

To this end, we parametrize the Minkowskian supermanifold $\mathcal{SM}^{(3|4)}$ with three bosonic coordinates $x^a= (z, \bar{z}, \tau)$ $(a=1,2,3)$ and four real fermionic coordinates arranged into two Majorana $3d$ spinors $\theta_i^\alpha$ $(i=1,2 \ \text{and} \  \alpha=\pm)$. The corresponding super-vielbeins are
\begin{equation}   
\label{eq:3Dvielbeins}
V^a=dx^a+\theta_i\gamma^ad\theta_i~~~~~,~~~~~~   \psi_i^\alpha = d\theta_i^\alpha
\end{equation}
which are invariant under rigid supersymmetry transformations.
Their corresponding dual vector fields are given by 
\begin{equation}
    \partial_a ~~~,~~~D^i_\alpha=\partial^i_\alpha-(\theta^i\gamma^a)_\alpha \partial_a
\end{equation}
and they satisfy the following algebra
\begin{equation}
  \{D^i_\alpha,D^j_\beta\}=-2\delta^{ij}\gamma^a_{\alpha\beta}\partial_a
\end{equation}
In this basis, the differential operator can be written as 
\begin{equation}
    d=V^a\partial_a+\psi^\alpha_iD^i_\alpha
\end{equation}
while the supercharges implementing supersymmetry transformations are
\begin{equation}
Q^i_\alpha=\partial^i_\alpha+(\theta^i\gamma^a)_\alpha \partial_a
\end{equation}
 satisfying the following algebra
\begin{equation}
  \{Q^i_\alpha,Q^j_\beta\}=2\delta^{ij}\gamma^a_{\alpha\beta}\partial_a ~~~,~~~  \{Q^i_\alpha,D^j_\beta\}=0.
\end{equation}
\\ 
As a consistency check, one can reduce the $3d$ $\mathcal{N}=2$ superalgebra to the coordinates $(z,\bar z,\theta_1^+,\theta_2^-)$ (setting $\tau = 0$ and $\theta^-_1 = \theta^+_2 = 0$) and see that it reproduces the $\mathcal N=(1,1)$ superalgebra of appendix \ref{appA}, upon identifying $\theta\equiv\theta_1^+,\bar\theta\equiv \theta_2^-$. This is an explicit realization of supersymmetry breaking at the boundary.

Referring to the field content of the bulk action $S_{\rm sBF}$ in \eqref{susycompacglobal}, the most 
general expansions of $(1|0)$ and $(1|4)$ forms on the basis of super-vielbeins \eqref{eq:3Dvielbeins} read
\begin{equation}\begin{split}\label{FormcompA}
\mathcal{A}_1&=\left(\tilde A_aV^a\!\!+\tilde A^\alpha_{i,ab}V^aV^b\iota^i_\alpha\!\!+\tilde A_{(ij)}^{(\alpha \beta)}V^3\iota^i_\alpha\iota^j_\beta\right)\delta^4(\psi)\\
    \mathcal{A}_2&= A_aV^a + A^i_\alpha \psi_i^\alpha
\end{split}
\end{equation}
Using the conventional constraints, one can write the field strength $d\mathcal{A}_2$ as 
\begin{equation}\label{A2curvature}
    d\mathcal{A}_2=F_{ab}V^aV^b+(W^i\gamma_a)_\alpha V^a\psi_i^\alpha 
    \end{equation}
    where $F_{ab}= \partial_{[a}A_{b]}$ and $(W^i\gamma_a)_\alpha=\partial_a A^i_\alpha-D^i_\alpha A_a$.

Inserting these expansions in the bulk action \eqref{susycompacglobal}, it is easy to work out that $S_{\rm sBF}$ has the following form
\begin{equation} 
S_{\text{sBF}} =\frac{1}{4\pi}\int_{\mathcal{SM}^{(3|4)}}\mathcal{L}_{\text{sBF}}(A,\tilde A) \, V^3\delta^4(\psi) \, , \quad {\rm with} \quad \mathcal{L}_{\text{sBF}}(\tilde A,A)=
\epsilon^{abc}\left(\tilde A_aF_{bc} - \tilde A^i_{ab}\gamma_cW_i\right) 
\end{equation}

Now, if we compute its supersymmetry variation we obtain a boundary contribution of the form \eqref{bdryD}, where the contraction acts on the supervolume form $V^3 \delta^4(\psi)$. Exploiting the following identity
\begin{equation}
    \iota_\epsilon (V^3\delta^4(\psi))=\left( \epsilon_{abc}(\epsilon_i\gamma^a\theta_i) V^bV^c-V^3\epsilon_i^\alpha\iota^i_\alpha\right)\!\delta^4(\psi)
\end{equation}
we are eventually left with
\begin{equation} 
\label{AppD2}
\delta_\epsilon S_{\text{sBF}} =\frac{1}{2\pi}\int_{\partial\mathcal{SM}^{(3|4)}} \mathcal{L}_{\text{sBF}} (\tilde A|_\partial,A|_\partial) \, (\epsilon_i\gamma^3\theta_i) V^2\delta^4(\psi)
\end{equation}
Here the notation  $\tilde A|_\partial$ and $A|_\partial$ stands schematically for the components of $\mathcal{A}_1$ and $\mathcal{A}_2$ at the boundary $\tau=0$, according to the boundary conditions in \eqref{BC gauge}. 

To work them out explicitly, we have to specify the choice of the PCO in \eqref{BC gauge}. 
It should encode the information about half of the supersymmetries that get broken by the boundary. In other words, the PCO should manifestly break super-translations along half of the fermionic directions. 

In order to preserve $\mathcal{N}=(1,1)$ supersymmetry at the boundary, without loss of generality we can choose 
\begin{equation}
\mathbb{Y}^{(0|2)}=\theta_1^-\theta_2^+\delta(\psi_1^-)\delta(\psi_2^+)
\end{equation}
and identify the other fermionic directions with the ones at the boundary, namely $(\theta_1^+, \theta_2^-) \equiv (\theta,\bar\theta)$. 

Recalling that the most general $(1|0)$ and $(1|2)$ forms on $\mathcal{SM}^{(2|2)}$ can be expanded respectively as
(now $a=1,2$)
\begin{equation}
\begin{split}
      B&=b_aV^a+ b_\psi \psi+\bar{b}_\psi \bar{\psi}\\   
     \widetilde B &=\widetilde b_aV^a\delta^2(\psi)+ \widetilde{b}_\psi V^2\iota\delta^2(\psi)+ \overline{\widetilde{b}}_\psi V^2\bar\iota\delta^2(\psi).
\end{split}
\end{equation}
and using the explicit expression for the PCO, the boundary conditions in \eqref{BC gauge} written in superfield components read  
\begin{eqnarray}\label{BC}
    \nonumber  A_a|_\partial&=& b_a ~~~~~~~~~~~\tilde A_a|_\partial= (\theta^-_1\theta^+_2) \widetilde b_a\\ \nonumber
     A_3|_\partial&=&0 ~~~~~~~~~~~~~\tilde A_3|_\partial=0 \\ 
A^1_+|_\partial&=& b_\psi~~~~~~~~
      (\tilde A_1)^+_{ab}|_\partial=\frac{1}{2}(\theta^-_1\theta^+_2)\epsilon_{ab} \, 
\widetilde{b}_\psi \nonumber\\
    A^2_-|_\partial&=&  \overline{b}_\psi ~~~~~~~~ 
(\tilde A_2)^-_{ab}|_\partial=\frac{1}{2}(\theta^-_1\theta^+_2)\epsilon_{ab}
\, \overline{\widetilde{b}}_\psi  \\
 A^1_-|_\partial&=& 0~~~~~~~~
      (\tilde A_1)^-_{ab}|_\partial=0 \nonumber \\
    A^2_+|_\partial&=&  0 ~~~~~~~~
   \nonumber
(\tilde A_2)^+_{ab}|_\partial=0 \\ \nonumber
 (\tilde A_i)^\alpha_{a3}|_\partial&=&0 ~~~~~~(\tilde A_{(ij)})^{\alpha\beta}|_\partial=0 
\end{eqnarray}
One can easily check that with such boundary conditions the expression $\mathcal{L}_{\text{sBF}}(A|_\partial,\tilde A|_\partial)$ is identically vanishing, therefore the symTFT is manifestly supersymmetric in ${\mathcal {SM}}^{(3|4)}$.

\bibliography{super.bib}

\end{document}